\documentclass[fleqn,10pt]{wlscirep}
\usepackage[utf8]{inputenc}
\usepackage[T1]{fontenc}
\usepackage{multirow}
\usepackage{lineno}
\usepackage[skip=3pt]{caption}
\AtBeginDocument{%
  \providecommand\BibTeX{{%
    \normalfont B\kern-0.5em{\scshape i\kern-0.25em b}\kern-0.8em\TeX}}}

\title{Using Large Language Models to Accelerate Communication for Users with Severe Motor Impairments}

\author[1,*]{Shanqing Cai}
\author[1]{Subhashini Venugopalan}
\author[1]{Katie Seaver}
\author[1]{Xiang Xiao}
\author[1]{Katrin Tomanek}
\author[1]{Sri Jalasutram}
\author[1]{Meredith Ringel Morris}
\author[1]{Shaun Kane}
\author[1]{Ajit Narayanan}
\author[1]{Robert L. MacDonald}
\author[2]{Emily Kornman}
\author[2]{Daniel Vance}
\author[2]{Blair Casey}
\author[2]{Steve M. Gleason}
\author[1]{Philip Q. Nelson}
\author[1,3]{Michael P. Brenner}
\affil[1]{Google, Mountain View, CA, USA}
\affil[2]{Team Gleason Foundation, New Orleans, LA, USA}
\affil[3]{School of Engineering and Applied Sciences, Harvard University, Cambridge, MA, USA}

\affil[*]{corresponding author(s): \newline Shanqing Cai
(cais@google.com)}

\begin{abstract}
Finding ways to accelerate text input for individuals with profound motor impairments has been a long-standing area of research. Closing the speed gap for augmentative and alternative communication (AAC) devices such as eye-tracking keyboards is important for improving the quality of life for such individuals. Recent advances in neural networks of natural language pose new opportunities for re-thinking strategies and user interfaces for enhanced text-entry for AAC users. In this paper, we present SpeakFaster, consisting of large language models (LLMs) and a co-designed user interface for text entry in a highly-abbreviated form, allowing saving 57\% more motor actions than traditional predictive keyboards in offline simulation. A pilot study with 19 non-AAC participants typing on a mobile device by hand demonstrated gains in motor savings in line with the offline simulation, while introducing relatively small effects on overall typing speed. Lab and field testing on two eye-gaze typing users with amyotrophic lateral sclerosis (ALS) demonstrated text-entry rates 29-60\% faster than traditional baselines, due to significant saving of expensive keystrokes achieved through phrase and word predictions from context-aware LLMs. These findings provide a strong foundation for further exploration of substantially-accelerated text communication for motor-impaired users and demonstrate a direction for applying LLMs to text-based user interfaces.

\end{abstract}
\begin{document}

\flushbottom
\maketitle

\thispagestyle{empty}

\section{Introduction}

The loss of efficient means of communication presents one of the greatest challenges to people living with amyotrophic lateral sclerosis (ALS). Absent typing or speaking, eye-gaze tracking is the most common human-computer interface. Coupled with an on-screen keyboard, gaze tracking allows users to enter text for speech output and electronic messaging, providing a means for augmentative and alternative communication (AAC)\cite{majaranta2002twenty}. However, gaze typing is slow (usually below 10 words per minute, WPM\cite{majaranta2002twenty, majaranta2007text}), creating a gap of more than an order of magnitude below typical speaking rates (up to 190 WPM in English\cite{yorkston1996sentence}). With recent developments in Brain-Computer Interfaces (BCI), typing via brain signals becomes possible and is allowing disabled users to type at increasingly higher rates through imagined speech or hand movements\cite{willett2021high, willett2023high},  however these methods require invasively-implanted electrodes in the cerebral cortex and hence are not as widely tested or adopted as eye-gaze typing.

The key bottleneck in gaze typing for disabled users is the high motor and temporal cost it takes to perform each keystroke\cite{cai2023speakfaster}. This bottleneck applies to BCI text entry systems that operate at the level of individual characters\cite{willett2021high, chen2015high} as well. One way to attack this bottleneck is to develop techniques to significantly reduce the number of keystrokes needed to enter text\cite{koester1994modeling, valencia2023less, kreiss2023practical}. For example, completion for partially-typed words and prediction of upcoming words could reduce the average number of keystrokes needed per word. However current keystroke saving techniques used in commercial text-input software (e.g., Gboard\cite{hard2018federated} and Smart Compose\cite{chen2019gmail}) are still insufficient for closing the communication gaps for impaired users, since the spans of correct predictions are short and the utilization of the predictions too infrequent to offset the cost of scanning predictions\cite{trnka2009user}. 

In the past few years, there has been a revolution in the development of Large Language Models (LLMs)\cite{thoppilan2022lamda, devlin2018bert, brown2020language, adiwardana2020towards}. LLMs are initially trained on simple objectives such as predicting the next text token over a very large text corpus, such as public web text and book content, but then can be nudged to have emergent capabilities that approach or exceed human-level performance on language-based tasks ranging from answering difficult mathematics problems\cite{lewkowycz2022solving} to carrying on open-ended conversations\cite{thoppilan2022lamda}. These LLMs can make predictions with a much longer context than traditional n-gram language models and thus offer the possibility of substantially accelerating character-based typing.  

In this study, we report a text-entry user interface (UI) called SpeakFaster that leverages fine-tuned LLMs to expand highly-abbreviated English text (word initials, supplemented by additional letters and words when necessary) into the desired full phrases at very high accuracy. To demonstrate the efficacy of this new system, we first used simulations to show that the accuracy of the SpeakFaster method is sufficiently high to significantly accelerate character-based typing. Second, we integrated the SpeakFaster UI with the eye-gaze keyboards of two users with ALS and demonstrated a significant enhancement in text-entry speed (29-60\%) and saving of motor actions compared to a baseline of conventional forward suggestions. On non-AAC participants the overall speed of text entry was similar to conventional smart keyboards while the motor action savings was still substantial. Our findings also indicate that users were able to learn to use the system with relatively little practice (about 20-50 phrases).

\section{LLMs and text-entry UI}

\begin{figure}
\centering
\includegraphics[width=\linewidth]{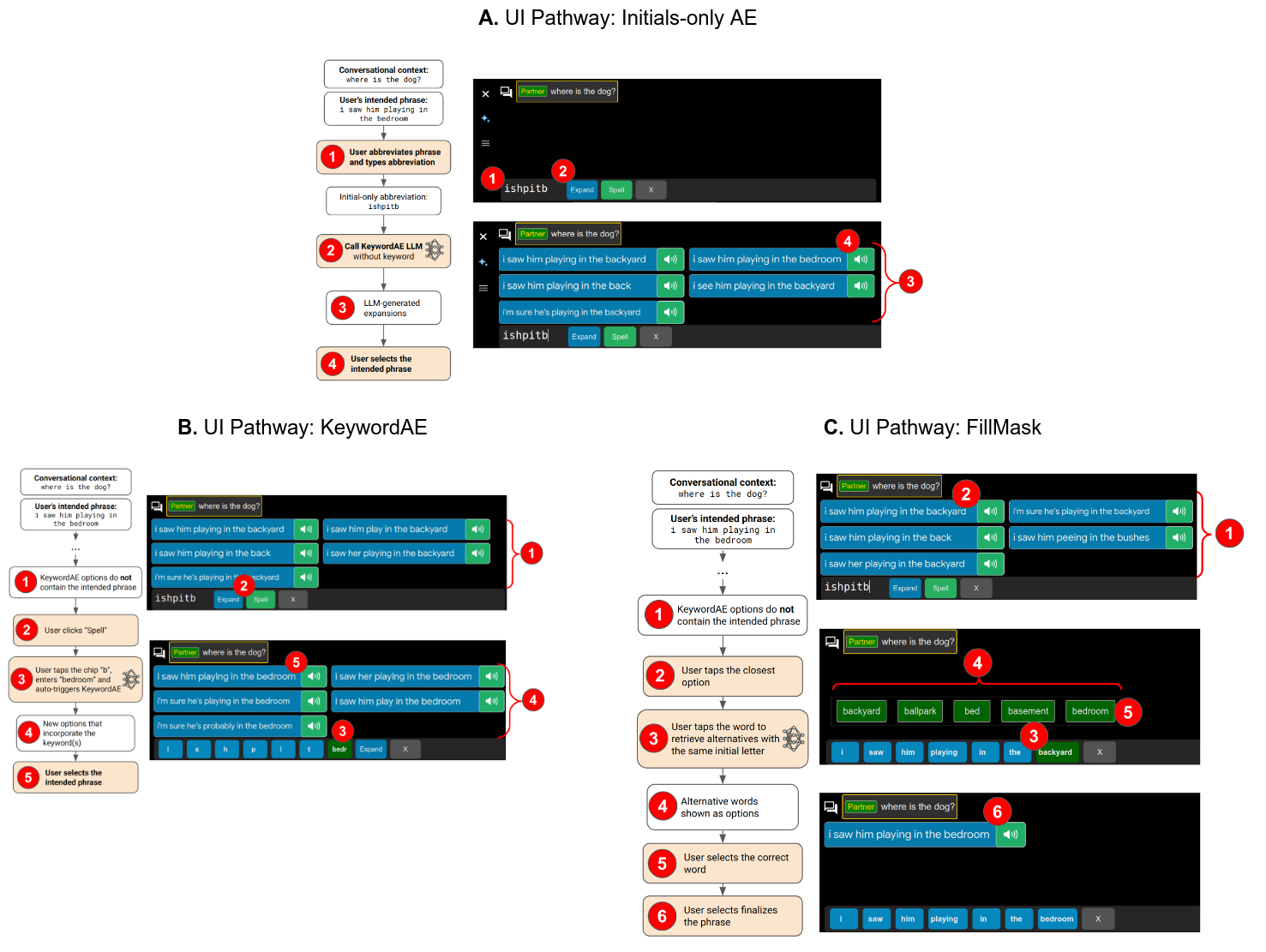}
\caption{Three interaction pathways of abbreviated text entry in the SpeakFaster UI subserved by two fine-tuned LLMs (KeywordAE and FillMask). Panel A shows the abbreviation expansion (AE) UI pathway with only word-initial letters. The “Speaker” button to the right of each candidate phrase (e.g., Label 4) allows the user to select and speak the correct phrase out via text-to-speech. Gaze-clicking the “Expand” button (Label 2) is optional since calls to the LLM are triggered automatically following the eye-gaze keystrokes. Panel B illustrates KeywordAE, which is an extension of initials-only AE that allows words that are spelled completely or incompletely to be mixed with initials in the original abbreviation, in an ordering that corresponds to the intended phrase. We refer to such fully or partially spelled words as “keywords”. They guide the UI towards the intended phrase.  Label 3 (“bedr” for “bedroom”) illustrates the support for partially-spelled keywords supported by KeywordAE v2, which differs from KeywordAE v1, where only fully-spelled keywords (e.g., “bedroom”) are supported. Panel C illustrates the FillMask UI pathway, another interaction flow which allows the user to find sensible replacements for an incorrect word that starts with the same letter. The gaze-driven on-screen keyboard is omitted from the screenshots in this figure.}
\label{fig:pathways}
\end{figure}

Word completion and next-word prediction based on n-gram language models\cite{trnka2009user, quinn2016cost} exploit the statistical dependence of a word on a small number of (typically up to four) preceding words. By contrast, LLMs are able to take advantage of broader context, including tens or hundreds of preceding words entered by the user and previous turns of the ongoing conversation. We have previously demonstrated\cite{cai2022context} that a fine-tuned 64-billion-parameter Google LaMDA\cite{thoppilan2022lamda} model can expand abbreviations of the word-initial form (e.g., “ishpitb”) into full phrases (e.g., “I saw him play in the bedroom”, Fig.~\ref{fig:pathways}A) at a top-5 exact-match accuracy of 48-77\% when provided with conversational context. Failures to find exact matches tend to occur on longer and more complex phrases. While encouraging, a practical solution needs to ensure that the user is able to type any arbitrary phrase in subsequent attempts in case of a failure in the initial abbreviation expansion, i.e., that the user will never run into a “dead end” in the UI.
  
We therefore developed a new UI and two underlying new fine-tuned LLMs as a complete, practical solution. LLM “KeywordAE” is capable of expanding abbreviations that mix initials with words that are fully or incompletely spelled out (Fig.~\ref{fig:pathways}B). The KeywordAE model is also capable of expanding initials-only abbreviations, and hence provides a superset of the capabilities of the fine-tuned LLM in Cai et al.\cite{cai2022context} LLM “FillMask” is capable of providing alternative words that begin with a given initial letter in the context of surrounding words (Fig.~\ref{fig:pathways}C). The two models models were each fine-tuned with approximately 1.8 million unique triplets of ${\{context, abbreviation, full phrase\}}$ synthesized from four public datasets of dialogues in English (See Supplementary Section~\ref{sec:supp_s1} for details of LLM fine-tuning and evaluation). 

To form conduits to the fine-tuned LLMs, we designed a UI with three pathways, namely initials-only AE, KeywordAE, and FillMask, to support a complete abbreviated text input experience (see the three panels of Fig.~\ref{fig:pathways}). The initials-only pathway is the common starting point of all phrase-entry workflows in the SpeakFaster UI. Among the three pathways, it involves the fewest keystrokes and clicks and alone suffices for short and predictable phrases. The user starts by typing an initialism that represents the intended phrase (Label 1 in Fig.~\ref{fig:pathways}A). The abbreviation is typed with a conventional soft keyboard or gaze-driven on-screen keyboard (e.g., Tobii ® Computer Control). As the user enters the abbreviation, the UI automatically triggers calls to the KeywordAE LLM after every keystroke. Each call returns the top-5 most likely options based on the conversational context and the abbreviation, which are rendered in the UI for the user to peruse and select (Label 3 in Fig.~\ref{fig:pathways}A). The number of options provided (5) is based on the screen size of gaze tablet devices and the point of "diminishing return" for motor saving rate from offline simulation results (see next section). If one of the candidate phrases matches the intended phrase, the user selects the phrase by clicking the “Speaker” button associated with it (Label 4 in Fig.~\ref{fig:pathways}A), which dispatches the phrase for text-to-speech output and ends the phrase entry workflow. The selected phrase ("I saw him playing in the bedroom" in this example) becomes a part of the conversational context for future turns.

If the intended phrase is not found via the initials-only pathway, however, two alternative UI pathways are available to assist the user in finding the intended phrase. One of the pathways is KeywordAE. The user gaze-clicks the "Spell" button (Label 2 in Fig.~\ref{fig:pathways}B), which turns the abbreviation in the input bar into gaze-clickable chips, one for each character of the initials-only abbreviation (e.g., bottom of Fig.~\ref{fig:pathways}B). The user selects a word to spell by gaze-clicking the corresponding chip. This turns the chip into an input box, in which the user types the word by using the on-screen keyboard (Label 3 in Fig.~\ref{fig:pathways}B). Subsequent calls to the LLM will contain the partially- or fully-spelled words in addition to the initials of the unspelled words. A call to the KeywordAE is triggered automatically after every keystroke. After each call, the UI renders the latest top-5 phrase expansion returned by the Keyword AE LLM (Label 4 in Fig.~\ref{fig:pathways}B). If the intended phrase is found, the user selects it by a gaze click of the speaker button as described before (Label 5 in Fig.~\ref{fig:pathways}B). We constructed two versions of KeywordAE models: KeywordAE v1 requires each keyword to be typed in full, while KeywordAE v2 allows a keyword to be typed incompletely (“bedroom” as “bedr”). Simulation results below show v2 leads to greater keystroke saving than v1.

The KeywordAE UI pathway is not limited to spelling out a single word. The UI allows the user to spell multiple words, which is necessary for longer and more unpredictable phrases (not shown in Fig.~\ref{fig:pathways}). In the unlikely case where the AE LLM predicts none of the words of the sentence correctly, the KeywordAE pathway reduces to spelling out all the words of the phrase.

FillMask is another pathway to recover from the failure to find the exact intended phrase. Unlike KeywordAE, FillMask only suits cases in which very few words (typically one word) of an expansion are incorrect (i.e., the phrase is a “near miss”). For instance, one of the phrases "I saw him play in the backyard" missed the intended phrase "I saw him play in the bedroom" by only one incorrect word ("backyard", Label 2 in Fig.~\ref{fig:pathways}C). The user clicks the near-miss phrase, which causes the words of the phrase to appear as chips in the input bar. Clicking the chip that corresponds to the incorrect word ("backyard", Label 3 in Fig.~\ref{fig:pathways}C) triggers a call to the FillMask LLM, the response from which contains alternative words that start with the same initial letter and fit the context formed by the other words of the sentence and by the previous turn(s) of the conversation. The user selects the correct word (Label 5) by clicking it and then clicking the speaker button to finalize the phrase entry (Label 6).

Although not shown by the example in Fig.~\ref{fig:pathways}C, the FillMask pathway allows the UI to replace multiple words (or do replacement multiple times in a given word slot) after the initial replacement. In rare cases where the FillMask LLM fails to provide the intended word, the user can fall back to typing the correct word in the input box by using the eye-gaze keyboard. 

As shown above, KeywordAE and FillMask are two alternative interaction modes for recovering from a failure to obtain the intended phrase via the initials-only pathway. The user’s decision on which pathway to choose should be determined by whether a near-miss option exists. This proposition is supported by simulation results in the next section. In the current study, the SpeakFaster UI allows the user to use the FillMask mode after using the KeywordAE mode, which is useful for finding the correct words in a hard-to-predict phrase. But entering the KeywordAE mode is not allowed after using the FillMask mode, because FillMask should be used only during the final phase of a phrase entry workflow, where all but one or two words of a candidate phrase are correct. These heuristics and design considerations of the UI pathways were made clear to the users through initial training and practice at the beginning of the user studies described below. The SpeakFaster UI is only one of many possible UI designs for supporting AAC text entry with LLMs\cite{valencia2023less, kreiss2023practical}. Its justification comes from prior studies on LLM’s capabilities in expanding abbreviations\cite{cai2022context}, its consistency with the conventional lookup-based abbreviation expansion in AAC\cite{demasco1994human}, and the empirical results from the user studies reported below.
\section{Simulation Results}

\begin{figure}
\centering
\includegraphics[width=0.9\linewidth]{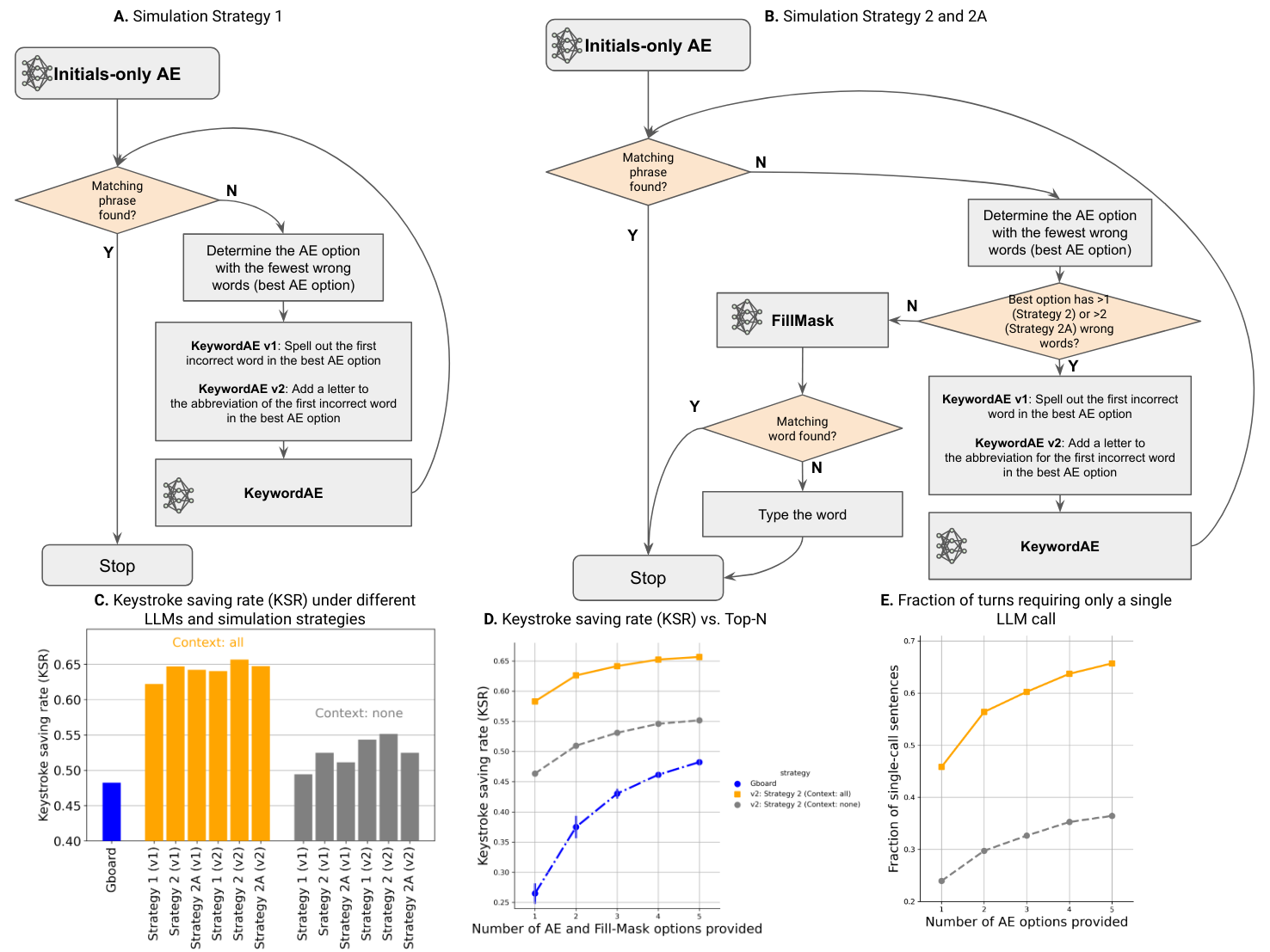}
\caption{Simulation of phrase entry assisted by the Keyword Abbreviation Expansion (KeywordAE) and FillMask LLMs in SpeakFaster. \textbf{A:} Simulation Strategy 1: AE with only initials is followed by KeywordAE if the initials-only AE doesn't return the desired phrase. Keyword AE in v1 iteratively spells out the first (leftmost) words in the best-matching phrase option until the desired phrase is found. KeywordAE in v2 iteratively appends letters to the first (leftmost) incorrect word in the best-matching phrase option. \textbf{B:} Simulation Strategy 2: same as Strategy 1, except that FillMask is used whenever only one incorrect word remains in the best-matching phrase. \textbf{C:} Keystroke saving rates (KSRs) from the simulation strategies, compared with a forward-prediction baseline from Gboard (blue bar). The orange bars show the KSRs when conversational context is utilized, while the blue bars show the KSRs without utilizing conversational context. All data in this plot are based on 5-best options from the KeywordAE and FillMask LLMs. \textbf{D:} the KSRs from Strategies 2, plotted as a function of the number of LLM options, in comparison with Gboard forward prediction. E: Fraction of dialogue turns that could be successfully entered with initials-only AE call, as a function of the number of options provided and the availability of conversational context. All results in this figure are from simulations on the dialogue turns for which the sentence length was 10 or shorter (counting words and mid-sentence punctuation) from the 280 dialogues in the test split of the Turk Dialogues (Corrected) dataset\cite{vertanen2017towards, cai2022context}.}
\label{fig:sim_results}
\end{figure}

To measure the approximate upper bound of the motor-action savings in this new text-entry UI, we ran simulation on the test split of a corrected version of the Turk Dialogues (TDC) corpus\cite{vertanen2017towards, cai2022context}. To simulate an ideal user’s actions in the SpeakFaster UI when entering the text for a dialogue turn, we first invoked AE without keywords (Fig.~\ref{fig:pathways}A). If the matching phrase was found, the phrase was selected and the simulation ended. If no matching phrase was found, however, we tested three interaction strategies. The first strategy (Strategy 1) invoked the KeywordAE (Fig.~\ref{fig:pathways}B) iteratively by spelling more of the words out, until the matching phrase was found. The second strategy (Strategy 2) was identical to Strategy 1, except FillMask (Fig.~\ref{fig:pathways}C) was used in lieu of KeywordAE whenever there remained only a single incorrect word in the best-matching phrase candidate. The flowcharts for Strategies 1 and 2 are shown in Fig.~\ref{fig:sim_results}A and ~\ref{fig:sim_results}B, respectively. The third strategy, referred to as Strategy 2A, was a variant of Strategy 2. It utilized FillMask more aggressively, i.e., as soon as two or fewer incorrect words remained in the best option. In all three strategies, KeywordAE was invoked incrementally by spelling out more words in the best-matching candidate. This incremental spelling was implemented differently for the two versions of KeywordAE, due to differences in what abbreviations were supported. For KeywordAE v1, which supported only fully-spelled keywords, the simulation spelled out the first incorrect word in the best option; for v2, in which keywords could be partly spelled, the simulation added one additional letter at a time in the first incorrect word. To leverage the contextual understanding of the AE and FillMask LLMs, all the previous turns of a dialogue from the Turk Dialogues corpus were used for expanding abbreviations and finding alternative words, unless otherwise stated.

As a baseline for comparison and to emulate the traditional n-gram-based text prediction paradigm, we also ran simulations on the same corpus with an n-gram language model (LM,  Gboard’s finite state transducer\cite{van2019writing} trained on 164,000 unigrams and 1.3 million n-grams in US English) that supported word completion and prediction, modeling ideal user behavior that selects the intended word as soon as it became available among the top-n word completions or prediction options.

To quantify the number of motor actions, we broadened the definition of keystrokes to include not only keypresses on the keyboard but also UI actions required to use the novel features in SpeakFaster, including entering the KeywordAE mode (a click to the “Spell” button in Fig.~\ref{fig:pathways}B), specifying a word to spell for KeywordAE, entering the FillMask mode, selecting a word to replace through FillMask, and selecting phrase and word options returned by the LLMs. Similarly, the number of keystrokes in the Gboard simulation included selecting options from word completion and next-word prediction.

The result of the SpeakFaster simulations indicated a significant saving of motor actions compared to the baseline from Gboard’s forward predictions (Fig.~\ref{fig:sim_results}C). This held true for both KeywordAE v1 and v2. Under KeywordAE v2, given that SpeakFaster utilized all previous dialogue turns as the context and provided five options at each step (orange bars in Fig.~\ref{fig:sim_results}C), Strategy 1 and Strategy 2 led to the KSR values 0.640 and 0.657, respectively, significantly exceeding the Gboard KSR (0.482). These KSRs from the KeywordAE v2 model also beat the best KSR from KeywordAE v1 (0.647) by a small but noticeable margin, reflecting a benefit of allowing keywords to be partially spelled. The superior KSR of Strategy 2 relative to Strategy 1 indicates a benefit of augmenting KeywordAE with FillMask, which surfaced the correct word options with fewer motor actions required. However, the comparison between Strategy 2 and Strategy 2A shows that FillMask deteriorates motor saving rate if used too aggressively. Specifically, premature uses of FillMask, i.e., whenever two incorrect words remained (instead of one incorrect word as in Strategy 2), reduced the KSR from 0.657 to 0.647 (Fig.~\ref{fig:sim_results}C). Under KeywordAE v1, the KSR showed a similar trend in relation to utilization of FillMask under Strategies 2 and 2A (not shown in Fig.~\ref{fig:sim_results}C for clarity). In Fig.~\ref{fig:sim_results}C, the gray bars show that SpeakFaster outperformed the Gboard baseline in KSR even without utilizing the previous dialogue turns as context, although the KSR gains were significantly lower compared to if the context was utilized.

The results in Fig.~\ref{fig:sim_results}C are all based on providing 5-best phrase options from the KeywordAE and FillMask LLMs. To illustrate the effect of varying the number of LLM options, Fig.~\ref{fig:sim_results}D plots the KSRs against the number of LLM options. Similar to the trend from Gboard, KSRs in SpeakFaster increased monotonically with the number of options, but started to level off at approximately five, which forms the basis for our UI design decision of including 5-best options (Fig.~\ref{fig:pathways}). Fig.~\ref{fig:sim_results}E shows that when conversational context was made available to the KeywordAE LLM (either v1 or v2), approximately two-thirds of the dialogue turns in the test corpus could be found with only the initials-only AE call (i.e., a single LLM call). This fraction became approximately halved when the conversational context was unavailable, which again highlights the importance of conversational context to the predictive power of the LLMs.

The simulation results above show that the theoretical motor-action saving afforded by context-aware AE and FillMask surpassed that of the traditional forward prediction by 30-40\% (relative). This result builds on the previous AE LLM in Cai et al. (2022) and goes a step further by supporting abbreviations that include spelled words (KeywordAE) and suggesting alternative words (FillMask), which removes “dead ends”, thus allowing any arbitrary phrase to be entered. However, as shown by prior studies\cite{li2022c, koester1994modeling, trnka2009user, quinn2016cost}, the predictive power of motor action-saving features in text-entry UIs is often offset by the added visual and cognitive burden involved in using these features, besides human errors such as misclicks and misspellings. Our system additionally involved network latencies due to calls to LLMs running in the cloud (see Supplementary Section~\ref{sec:supp_s3}). Therefore, the practical performance of the LLM-powered abbreviation expansion-based text-entry paradigm in SpeakFaster must be tested with empirical user studies. To this end we conducted a controlled lab study on a group of users typing manually a mobile device as a pilot study of the novel text entry paradigm, followed by lab and field studies on two eye-gaze typing users with ALS. 

\section{User Studies}

We tested the SpeakFaster text-entry UI with two groups of users. First, a group of non-disabled users typed with their hands on a mobile touch-screen device running SpeakFaster powered by the KeywordAE v1 and FillMask LLMs. In a separate eye-gaze study, users with ALS who were experienced eye-gaze typers entered text by using eye-trackers integrated with SpeakFaster. The mobile user study served as a pilot for the eye-gaze study by proving the learnability and practicality of the novel text-entry UI. A common goal of the two studies was to understand the cognitive and temporal cost introduced by the SpeakFaster UI and how that affects the overall text-entry rate compared to a conventional baseline. To study this under different levels of spontaneity and authoring task load, our study protocol consisted of both a scripted phase and an unscripted phase. 

The scripted phase consisted of 10 dialogs from the test-split of the Turk Dialogues Corrected (TDC) corpus\cite{vertanen2017towards, cai2022context}. Each dialogue is a two-person conversation with a total of six turns, three uttered by each person. Our user study participant played the role of one of the persons in the conversation, and the to-be-entered text was displayed to them close to their typing area. In the unscripted phase, the user engaged in a set of five six-turn dialogues with the experimenter where only the starting question was predetermined, and the rest was spontaneous but required the user to keep each conversation turn under ten words and not include any personal information. The experimenter started with an open-ended question such as “What kind of music do you listen to?” (see Section~\ref{sec:supp_s4} for the full list) and then the user would reply. The experimenter and user then followed up in alternating turns, until a total of six turns was reached for each dialogue. In both the scripted and unscripted phases, for each block of five dialogues, the first and last dialogs formed the baseline, in which the user typed with the regular keyboard i.e. either Gboard in the mobile user study or the Tobii eye-gaze keyboard in the eye-gaze users study, utilizing word suggestions provided by the default keyboard at will. In the other three dialogs, the user entered text by using the SpeakFaster UI, starting out with the initials-only abbreviation scheme and using any of the pathways to either spell out words (KeywordAE) or select replacements (FillMask) as per their preference until they were satisfied.

Prior to the data collection portion of the lab study, each user watched a video demonstration of the SpeakFaster system and took a short practice run to familiarize themselves with the SpeakFaster UI and the abbreviation expansion text entry paradigm. Each mobile user study participant practiced a minimum of five practice dialogs. The content of these practice dialogs were different from the ones used in the subsequent study blocks. Prior to the unscripted phase, the users were also familiarized with the unscripted test condition through practicing a single dialog. The eye-gaze user practiced for four hours over two days and this is described more in the Section~\ref{sec:supp_s4}.

\subsection{Text-entry rate is comparable between SpeakFaster and baseline for mobile users}

In order to study abbreviated text entry under different degrees of motor cost, the 19 participants who provided informed consent were randomly assigned to two typing-posture groups in the mobile user study. Nine users were assigned to the one-finger group and instructed to type with only the index of their dominant hand (right hand for all these users). The remaining ten users were assigned to the no-constraint group and were given no limitation related to typing posture. They all operated with both hands during the experiment, with varied posture details in which fingers or thumbs were used. 

In the scripted portion of the user study, no significant difference was observed in the accuracy of text entry between SpeakFaster and the Gboard baseline. The average word error rates (WERs) of the OneFinger group were 1.55\% and 2.53\% under the baseline and SpeakFaster conditions, respectively. For the NoConstraint group, the respective average WER were 3.96\% and 2.89\%. A two-way linear mixed model (Posture $\times$ UI) on the WERs showed no significant main effect by Posture (z=-1.758, p=0.079) or UI (z=0.079, p=0.250). Nor was there a significant interaction in WER between Posture and UI (z=1.516, p=0.129).

\begin{figure}
\centering
\includegraphics[width=\linewidth]{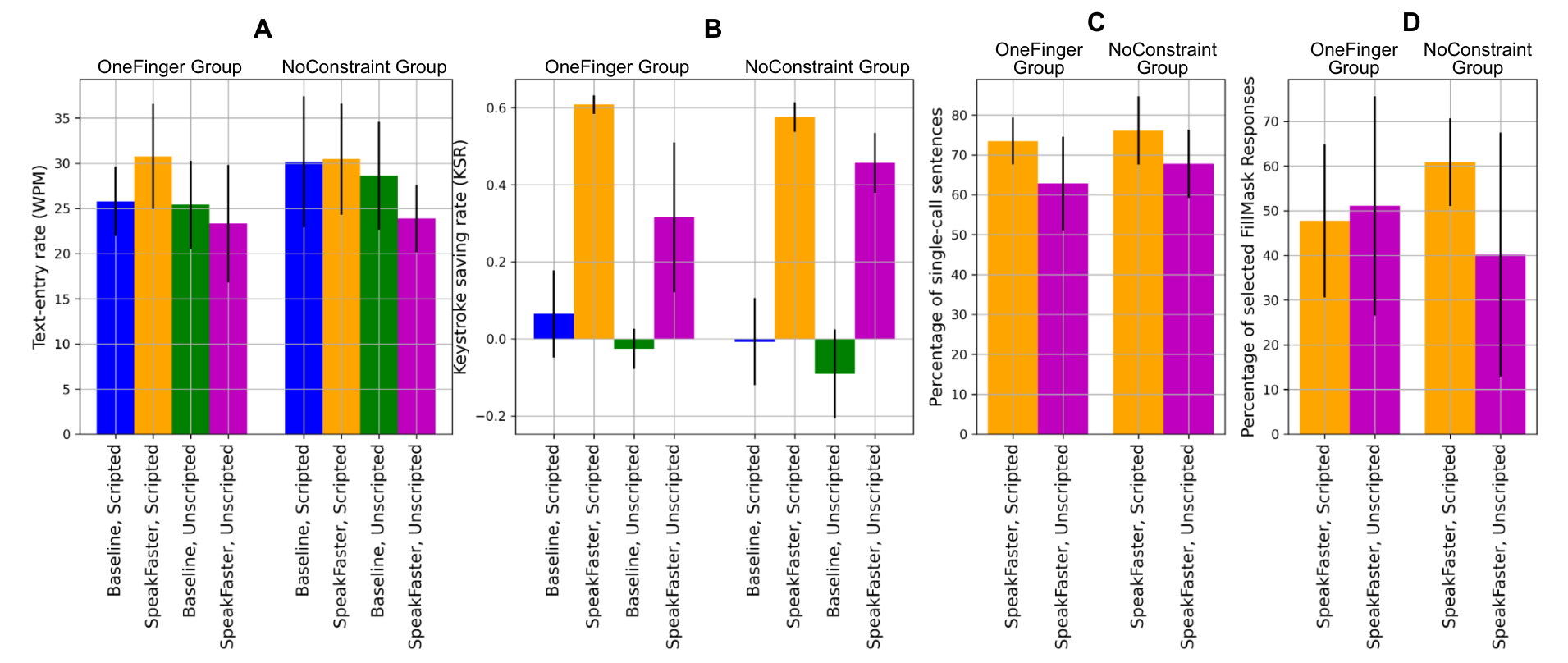}
\caption{Text-entry rate (Panel A), the keystroke-saving rate (KSR, Panel B), the percentage of sentences that involved only a single AE server call (Panel C), and the percentage of FillMask responses that contain options selected by users (Panel D) from the mobile study. In each plot, the two groups of bars correspond to the two posture groups (OneFinger and NoConstraint), as labeled at the top. The error bars in these plots show 95\% confidence intervals (CIs) of the mean. Overall, the text entry speed (WPM) using SpeakFaster was not significantly different from the baseline despite introducing significant keystroke savings.}
\label{fig:text_rate_mobile}
\end{figure}

The effect on text-entry rate by the LLM-powered SpeakFaster UI showed an intricate mixed pattern. While the text-entry rate saw increases on average under the SpeakFaster UI relative to the baseline (Gboard) UI during the scripted dialogs, the average rate showed a decrease when the users engaged in unscripted dialogs. Analysis by a linear mixed model did not reveal a significant main effect by UI (z=-0.959, p=0.338). However, a significant two-way interaction was found between UI and DialogType (z=2.933, p=0.003). Post hoc paired t-test confirmed a significant difference in the SpeakFaster-induced changes in the text-entry rate (relative to baseline) between the scripted and unscripted dialogs (t=-4.85, p=0.00013). Specifically, while SpeakFaster increased the average rate by 2.510$\pm$3.024 WPM (95\% CI of mean, relative: 13.0\%$\pm$24.5\%) under the scripted dialogs, it decreased the average rate by 3.494$\pm$3.294 (relative: 10.2\%$\pm$25.0\%) under the unscripted ones. The three-way linear mixed model did not reveal any other significant main effects or interactions.

\subsection{SpeakFaster enables significant keystrokes savings}

While the effect of SpeakFaster on the text-entry rate exhibited a complex pattern of interactions, showing an absence of overall significant change from the baseline, the keystroke-saving rate (KSR) was affected in a clearcut and pronounced manner by SpeakFaster (Fig.~\ref{fig:text_rate_mobile}B). The same three-way linear mixed model, when applied on the KSR as the dependent variable, revealed a significant main effect by UI (z=10.317, p=$6\times10^{-25}$). The linear mixed model revealed no other significant main effects or interactions. Relative to the Gboard baseline, the SpeakFaster UI paradigm led to a large and significant increase in KSR for both the scripted dialogs (+0.564$\pm$0.080 abs., p=$8.0\times10^{-11}$) and the unscripted ones (+0.450$\pm$0.114 abs., $p=5.5\times10^{-7}$). 

Panel C of Fig.~\ref{fig:text_rate_mobile} shows the percentage of dialog turn in which the user successfully entered the sentence by using only the initials-only AE call, i.e., without spelling out words in the abbreviation or using FillMask. As the orange bars show, the percentages were on par with the results from the simulation in the scripted dialogs (c.f. Fig.~\ref{fig:sim_results}E). The percentages of sentences that succeeded with a single AE call were lower for unscripted dialogs (magenta bars, 0.65 on average), reflecting the slight domain mismatch in the unscripted text content from the scripted ones that the AE and FillMask models were trained on.

\subsection{Simulation accurately predicts users keystroke savings}

\begin{figure}
\centering
\vspace{-1.0cm}
\includegraphics[width=0.36\linewidth,height=9.5cm]{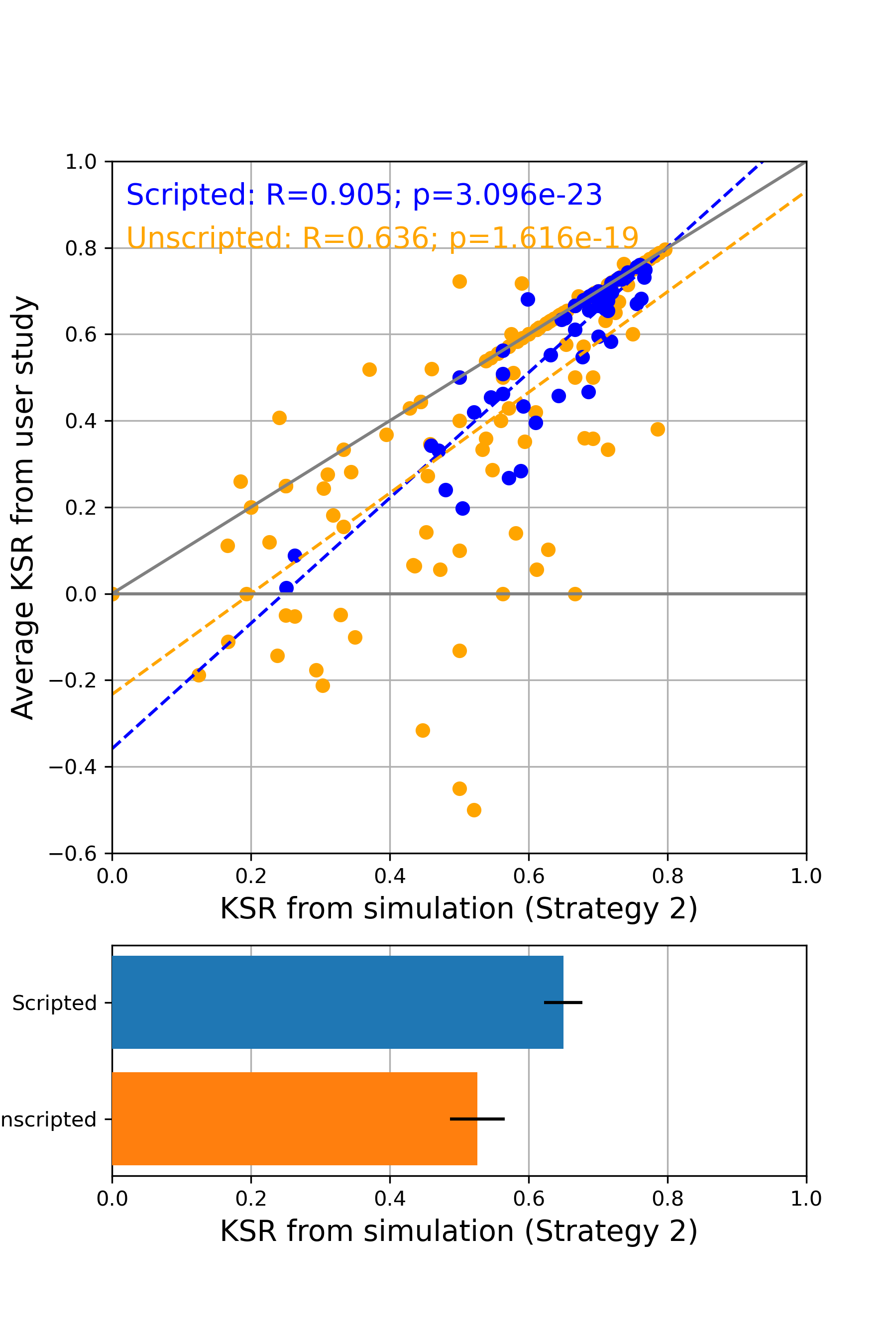}
\caption{The top panel shows correlations between the average KSRs observed in the users (y-axis) and the KSRs from offline simulation (x-axis), on a turn-by-turn basis. The blue and orange dots correspond to the scripted and unscripted dialogs, respectively. The simulation results are based on KeywordAE model v1, Strategy 2, five predictions, and using all previous turns of a dialog as the context. Each dot corresponds to a turn from one of the dialogs used as the text materials in the lab study. The bottom panel shows a comparison of the simulated KSR values from the scripted (blue) and unscripted (orange) dialog turns. The KSRs from the scripted turns were significantly higher than those from the unscripted ones (unpaired t-test: t=2.26, p=0.025).}
\label{fig:sim_proxy_keystrokes}
\vspace{-0.5cm}
\end{figure}

The KSR values observed from the users in the lab study could be predicted with a considerable accuracy by the simulation results. The blue dots in the top panel of Fig.~\ref{fig:sim_proxy_keystrokes} show a significant positive correlation between the average KSR values from all users and the simulated ones on a turn-by-turn basis among the scripted dialogs (Pearson's correlation: R=0.905, p=$3.096\times10^{-23}$). The unscripted dialogs (orange dots) also exhibited a significant correlation between the simulated and observed KSRs (R=0.636, p=$1.616\times10^{-19}$). However, it can be seen that the users' average performance did not fully realize the motor-saving potentials predicted by the offline simulations, as most data points in Fig.~\ref{fig:sim_proxy_keystrokes} fall below the line of equality (the solid diagonal line), potentially reflecting human errors such as typos and mis-operations, as well as the actions to recover from them. The degree to which the user behavior underperformed the simulation results was greater during the unscripted dialogs than the scripted ones. For example, several unscripted dialog turns showed negative KSRs, despite the fact that offline simulation predicted positive KSRs based on the final committed sentences. This reflects greater revisions due to human errors and change-of-mind when the users operate under the dual cognitive load of formulating a dialogue response and operating the SpeakFaster UI in order to enter the response.

The bottom panel of Fig.~\ref{fig:sim_proxy_keystrokes} shows that the simulated KSR values were significantly lower for the unscripted dialogs than the scripted dialogs (mean: 0.527 vs. 0.650, -18.99\% relative, unpaired t-test: p<0.05). This was likely due to a domain mismatch between the TDC dataset and the unscripted content composed by the users during the lab study. However, the fact that SpeakFaster significantly boosted KSR even for the unscripted dialogs (Fig.~\ref{fig:text_rate_mobile}) underlines the robustness of the motor saving paradigm against domain shifts.

\subsection{Temporal aspects of user interactions in SpeakFaster}

\begin{figure}[t]
\centering
\vspace{-1.0cm}
\includegraphics[width=0.9\linewidth,height=10cm]{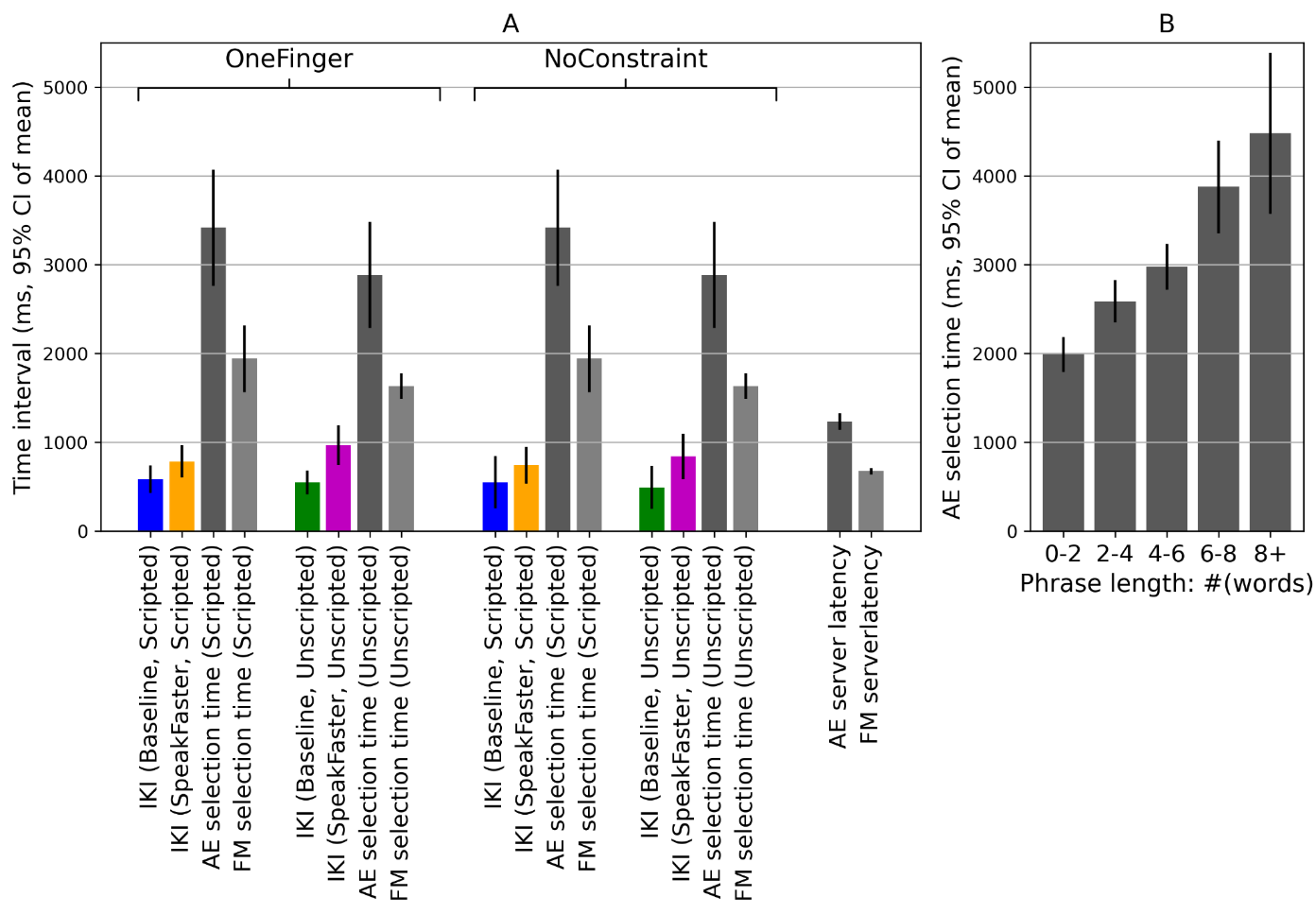}
\vspace{-0.5cm}
\caption{Temporal analyses of text entry in the SpeakFaster UI and the Gboard baseline in the mobile user study. \textbf{A:} the inter-keystroke intervals (IKIs), response times for selecting AE and FillMask (FM) options returned from the LLMs, and the latencies of calls to cloud-based LLMs. The bars are organized into five groups. The first two groups are from the OneFinger posture group, showing scripted and unscripted dialogs, respectively. Likewise, the third and fourth groups show the results from the NoConstraint posture group. The fifth group shows LLM latencies. \textbf{B:} the response time for AE option selection showed a clear increasing relation with the length of the underlying phrase, measured as number of words defined as the character length of the phrase divided by 5.}
\label{fig:temporal_analyses}
\vspace{-0.5cm}
\end{figure}

Figure~\ref{fig:temporal_analyses} shows the temporal aspects of how the mobile users interacted with the SpeakFaster UI. An inter-keystroke interval (IKI) is defined as the time interval between two consecutive keystrokes issued directly by the user using the soft keyboard (Gboard). IKIs exclude non-keystroke actions and the keystrokes automatically applied due to the user selecting options for word completion and predictions. The IKIs were significantly longer under the SpeakFaster UI than the baseline UI, showing that it was slower on average for the users to plan and perform the keystrokes when typing the characters of the abbreviations and performing the keystrokes required for subsequent spelling of words than typing words in the familiar, sequential fashion ( main effect by UI: z=3.671, p=$2.4\times10^{-4}$). The linear mixed model also identified a significant UI$\times$DialogType interaction (z=2.303, p=0.021). A post hoc t-test confirmed that the increase in the IKI was greater under the unscripted dialogs (+381.1 ms abs., 87.1\% rel.) than under the scripted ones (+194.8 ms abs. 51.7\% rel., p=0.0067). Similar to the observations related to KSRs above, this differential effect of SpeakFaster on IKI increase may be interpreted as the greater cognitive load under the dual task of composing a free-form dialog response and abbreviating the response in the SpeakFaster’s abbreviation regime.

The dark and light gray bars in Panel A of Fig.~\ref{fig:temporal_analyses} show the temporal metrics related to using the LLMs for the AE (including KeywordAE) and FillMask workflows, respectively. Compared with the latencies of the calls to the cloud-based LLMs (two rightmost bars in Panel A), the users’ average response times in selecting the AE-suggested phrases and FillMask-suggested words were 2-3 times longer. These response times not only exceeded the LLM latencies (by 2-3 times), but were also longer than the average IKIs by 3-6 times, indicating that they were a significant component of the total time it took to use the SpeakFaster UI. 

The average response times for AE were approximately twice as long as those for FM. This is attributable to the fact that the AE options are multi-word phrases while the FillMask options are single words, which highlights an additional benefit of the word-replacement interaction. As Fig.~\ref{fig:temporal_analyses}B shows, the AE response time showed a strong positive correlation with the length of the phrase (Spearman's $\rho$=0.428, p=$2.5\times10^{-20}$). While selecting phrase options of length two words or shorter took only approximately 2,000 ms on average, selecting phrases eight words or longer took more than twice as long.

The results of the mobile user study showed that the new LLM-based SpeakFaster text entry UI led users to achieve savings in motor actions including keystrokes up to 50 percentage points (absolute) higher than the conventional way of mobile text entry. In terms of speed, the results were mixed. While in the scripted dialogue condition the users achieved an average of 13\% speedup, they showed a 10\% slowdown under the unscripted condition, reflecting an interplay between the added cognitive load of using the new UI and that required to compose a spontaneous text message. Timing analysis revealed that reviewing the phrase options from the KeywordAE LLM took about 3-6 times the average IKI; it took relatively less time to review the word options in FillMask, but the review cost was still significant (2-3 times as long as the average IKI). These timing findings highlight an important trade-off between the cost of reviewing LLM outputs and the savings in motor actions. Mobile tap-typing IKIs were relatively short ($\approx$500 ms, Fig.~\ref{fig:temporal_analyses}A), which may mask the benefit due to the predictive power of the LLMs. However, in eye-gaze typing, the IKIs can be significantly longer, owing to the dwell time and the gaze travel time between keys. These considerations indicate stronger potentials for acceleration in eye-gaze text entry than mobile text entry. With this hypothesis, we proceeded to study eye-gaze users’ interaction with SpeakFaster.

\subsection{Eye-Gaze Users Study}

The two study participants were both adult males diagnosed with ALS (17 and 11 years prior to the study, respectively) and provided informed consent before participating in this study. Both users were native speakers of American English and experienced eye-gaze typists who communicate daily through eye trackers and associated keyboards. At the time of the study, both participants were quadriplegic, unable to speak, while retaining cognitive abilities within normal limits. Their eye movements remained functional and cognitive abilities were within normal limits. They were experienced with the Tobii ® eye-gaze on-screen keyboard and its n-gram word competition and prediction features similar to Gboard. The first participant engaged in a controlled lab study (LP1) and the second in a field deployment to gather more naturalistic data (FP1). 

\begin{figure}[t]
\centering
\includegraphics[width=0.6\linewidth]{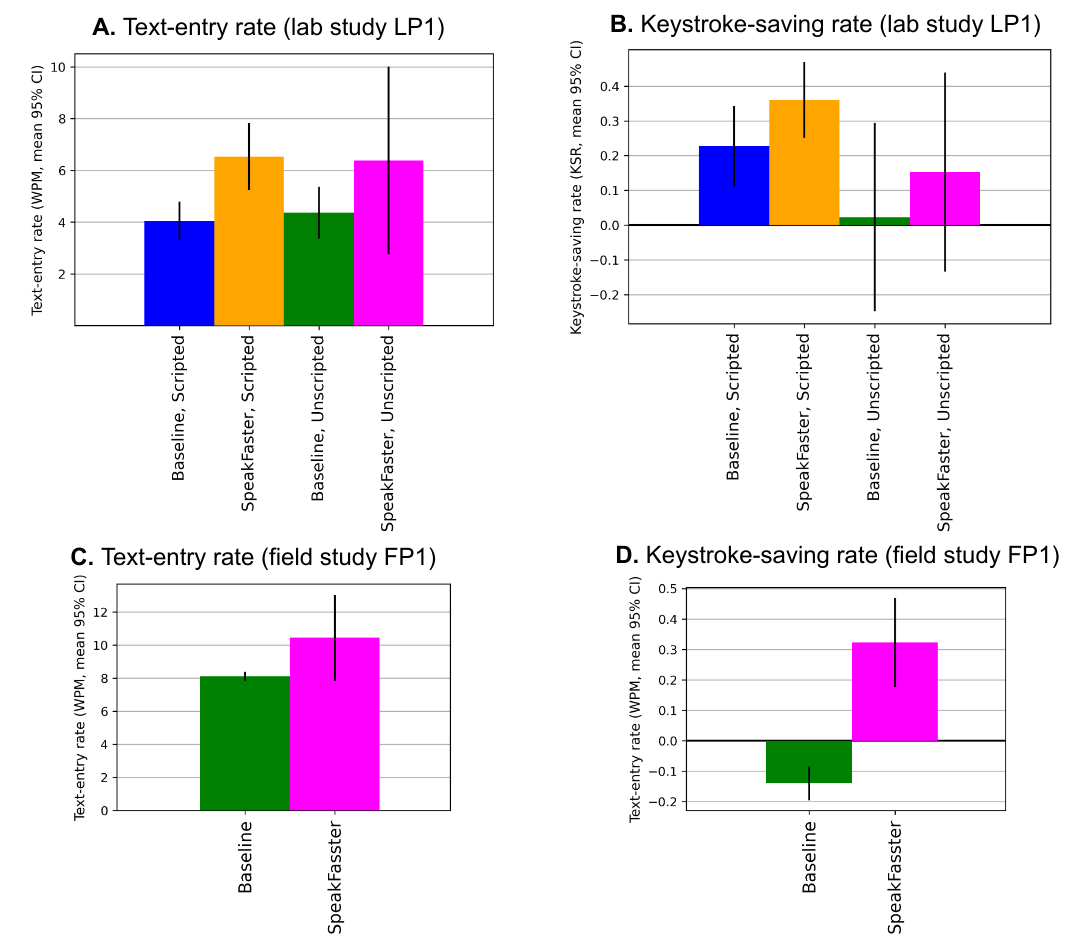}
\caption{Eye-gaze users' text-entry rate and KSR under the SpeakFaster testing condition and the baseline of Tobii on-screen keyboard. Panel \textbf{A} and \textbf{B} show the comparisons of the text-entry speed and keystroke saving rate (KSR) between SpeakFaster’s AE UI and a forward prediction baseline based on the Tobii eye-gaze keyboard in the lab study participant LP1. Panel \textbf{C} and \textbf{D} compares the text-entry rate and KSR of the field study participant FP1 while using SpeakFaster versus a Tobii keyboard baseline, for both scripted dialogues and unscripted dialogues in a controlled lab study. Error bars show 95\% confidence interval (CI) of the mean. Note the different y scales between the panels.}
\label{fig:eye_gaze_results}
\end{figure}

In the controlled lab study, our participant (LP1) followed a structured study protocol consisting of a scripted part followed by an unscripted part identical to that of the mobile user study described above. However, LP1 used the KeywordAE v2 model, which supported partially-spelled keywords and triggered LLM calls for initials-only AE and KeywordAE after every keystroke, eliminating the need for manual trigger of AE LLM calls as in the mobile user study. Prior to the data collections, the user practiced the SpeakFaster text entry paradigm under the direction of the experimenter for a total of 4.1 hours over two separate days before the lab study, for which the learning curve can be found in Supplementary section~\ref{sec:supp_s4}.

\subsection{SpeakFaster enables significant keystrokes savings and words per minute speed-up for eye-gaze users}
Figure~\ref{fig:eye_gaze_results}A compares the mean text-entry rates in words per minute (WPM) between the SpeakFaster paradigm with the Tobii keyboard baseline. Averaged over the scripted dialogues, the user achieved an average text-entry speed of 6.54 WPM while using SpeakFaster, which exceeded the baseline typing speed (4.05 WPM) by 61.3\% (two-sample t-test: $t_{28}$ = 2.76, p = 0.011). A similar rate enhancement occurred for the unscripted dialogues (SpeakFaster: 6.38 WPM, baseline: 4.37 WPM, 46.4\% increase), although the difference did not reach significance (t-test: $t_{13}$ = 0.818, p = 0.43). In addition to increased speed, a significant increase in the KSR was also observed for the user with SpeakFaster for both the scripted and unscripted dialogues. Again, statistical significance was reached only for the scripted dialogues (0.360 vs. the baseline of 0.227, rank-sum test: $\rho$ = 1.97, p = 0.049, Fig.~\ref{fig:text_rate_mobile}B). 

In the lab study, 77.8\% of the scripted dialogues and 66.7\% of the unscripted ones required only a single initials-only AE LLM call. For the trials in which LP1 used FillMask, the LLM predicted the correct words (as determined by user selection) 58.3\% of the time for the scripted dialogues and 21.4\% of the time for the unscripted ones. Despite achieving success a majority of the time (58.3\%), the FillMask LLM’s success rate observed on LP1 on the scripted dialogues is lower than the success rate predicted from offline simulation (70.7\%), indicating that the user occasionally failed to choose the correct words when they appeared. The fact that the success rate of the initial AE and FillMask was lower for unscripted dialogues reflects the domain mismatch between the user’s personal vocabulary and the model’s training corpus, highlighting personalization of the model as a useful future direction. 

To study the efficacy of the SpeakFaster paradigm under more natural usage, we conducted a field study with another eye-gaze user (FP1). As we reported previously 15, FP1 showed an average eye-gaze text-entry speed of 8.1$\pm$0.26 WPM (95\% confidence interval) over a period of six months of measurement in his real-life communication with close relatives and caregivers. This baseline speed of gaze typing is based on 856 utterances typed with the Tobii ® Windows Control eye-gaze keyboard with a PCEye Mini IS4 eye tracker. FP1 also engaged in test dialogues with an experimenter by using the SpeakFaster UI based on KeywordAE v1 and manual triggering of LLM calls for AE. Over the 27 phrases entered with SpeakFaster on six different days, the user achieved an average speed of 10.4$\pm$2.6 WPM, which is 28.8\% faster than the daily baseline (two-sample t-test: $t_{881}$ = 2.97, p = $3.1\times10^{-3}$, Fig.~\ref{fig:eye_gaze_results}C). Accompanying this increase in the average speed of text entry was an increase of keystroke saving rate (KSR) from -0.14 to 0.32 (rank-sum test: $\rho$ = 4.37, p = $1.21\times10^{-5}$, Fig.~\ref{fig:eye_gaze_results}D).

\subsection{Motor savings outweigh cognitive overhead for eye-gaze typing}

\begin{figure}[t]
\vspace{-.6cm}
\centering
\includegraphics[width=\linewidth]{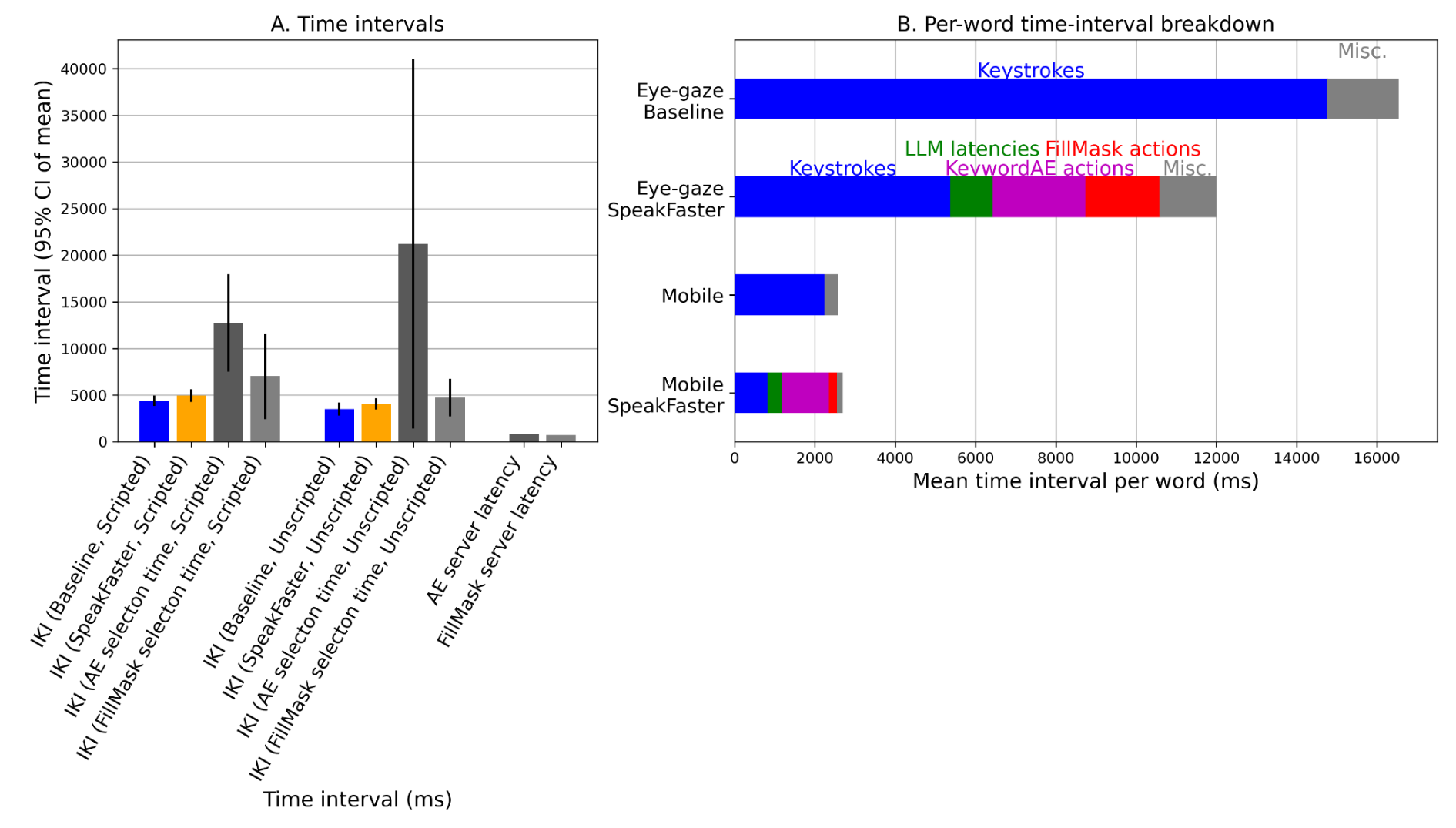}
\caption{Comparison of inter-keypress intervals, server latencies, and user-selection reaction times observed in the lab study on user LP1. Panel \textbf{A} compares the inter-keystroke  intervals (IKIs) of LP1’s eye-gaze keypresses, the time it took the user to select LLM-provided options for KeywordAE and FillMask, and the server latencies (including network latencies) for the LLM calls. Panel \textbf{B} shows breakdown of the amount of time spent on different tasks under the baseline and SpeakFaster typing conditions, observed in the eye-gaze user LP1 (two bars at the top) and the average over the 19 mobile-study users (two bars at the bottom). Only the data from the scripted typing condition are shown for clarity. The color code of the bar segments is omitted in the bottom two bars due to space limit, but is identical as that of the top two bars.}
\label{fig:cognitive_vs_motor}
\vspace{-0.4cm}
\end{figure}

The latencies for the AE calls were 843.0$\pm$55.4 ms and 832.7$\pm$120.4 ms (95\% CI of mean) for the scripted and unscripted dialogues, respectively (Fig.~\ref{fig:cognitive_vs_motor}A). The latencies of the FillMask calls were shorter (scripted: 617.9$\pm$41.8 ms; unscripted: 745.2$\pm$67.1 ms) than AE, due to its serving configuration that took advantage of the shorter output lengths (Supplementary section~\ref{sec:supp_s3}). These LLM-serving latencies were approximately four times shorter than the average eye-gaze IKIs measured on user LP1 (Fig.~\ref{fig:cognitive_vs_motor}A, blue and orange bars, 3,511 - 4,952 ms) and therefore had only a minor slowing effect on the text-entry rate of eye-gaze typers. In comparison, the time it took the user to select the correct AE responses was significantly longer (Fig.~\ref{fig:sim_proxy_keystrokes}B: scripted: 12,732$\pm$5,207 ms; unscripted: 21,225$\pm$19,807 ms), which was 3-6 times the average duration of a keypress, reflecting the significant cost for scanning the phrase options from the AE calls. By contrast, FillMask involved a much shorter (2-3x) candidate selection time than AE (Fig.~\ref{fig:sim_proxy_keystrokes}B: scripted: 7,032$\pm$4,584 ms; unscripted: 4,745$\pm$2,023 ms), reflecting the benefit of the FillMask interaction in providing shorter, single-word candidates, which reduced the scanning cost.

Compared with the average IKI in mobile users (Fig.~\ref{fig:temporal_analyses}), these IKIs from the eye-gaze typist shown in Fig.~\ref{fig:cognitive_vs_motor}A were 3 - 6 times longer. This provides insight as to why SpeakFaster leads to a significant speed up for eye-gaze typists in this study while introducing minimal changes among the participants in mobile user study described above. Specifically, Panel B of Fig.~\ref{fig:cognitive_vs_motor} shows a breakdown of the time intervals spent on several different types of actions during the baseline typing condition and SpeakFaster typing condition, as shown by different colors. The overhead of using SpeakFaster is broken down into the LLM latencies, the actions involved in using KeywordAE (entering the spelling mode, selecting a word to spell, as well as reviewing and selecting phrase options) and those involved in using FillMask (entering the FillMask mode, selecting the target word, as well as review and selecting word options). Among these subtypes of overhead, the LLM latencies were a relatively minor factor. The total overhead (1,735 ms/word) came close to the time spent on keystrokes (2,252 ms/word) for mobile users. A similar overhead is seen in the eye-gaze user (5,213 ms/word). However, this overhead was made up for by the dramatic reduction in time spent on keystrokes for the eye-gaze user (11,189 ms), leading to a reduction in the overall time on a per word basis. By contrast, the mobile users only showed a more modest average reduction of time spent on keystrokes (1,416 ms/word), which was insufficient to fully offset the increased overhead. This was largely due to the fact that the average IKIs were already much shorter during mobile typing than in eye-gaze typing.

In summary, the quantitative measurements of eye-gaze typing in both users found a significant advantage of the SpeakFaster UI of text input compared to the baseline of conventional eye-typing systems with forward-prediction. The evidence is seen both in a controlled lab setting and a field study in comparison with a real-life, long-term baseline.

\section{Discussion}

Abbreviation expansion of phrases in user-input history has long existed as a feature of eye-gaze AAC software, in addition to being seen in non-AAC applications such as text messaging,  and hence is familiar to users of such systems. SpeakFaster builds on this user knowledge, while drastically extending its breadth to open-ended phrases including previously-untyped phrases by using contextualized predictions from fine-tuned LLMs. Our demonstration of accelerated text entry focused on eye-gaze typing for motor-impaired users. However, the same LLM-based paradigm can be extended to other modalities of text entry in the AAC realm. For instance, SpeakFaster’s AE and FillMask design should serve to reduce the number of scans and selections required for text entry based on switch scanning, a type of user interface suitable for users who are unable to use eye tracking devices due to more pronounced motor impairments\cite{roark2010scanning}. Similarly, BCI-based text entry, including techniques based on motor imagery from surgical implants\cite{pandarinath2017high, willett2021high} and the noninvasive techniques based on visual-evoked potentials\cite{chen2015high} could benefit from the same greater predictive power of LLMs. Furthermore, for users without disabilities, constraints such as small screens and situational impairments\cite{sears2003computers, komninos2014text} may also render the temporal and motor cost of text entry high, in which case it will be interesting to explore the LLM-based techniques such as shown in the current study for motor savings and temporal acceleration. Such envisioned applications of the SpeakFaster paradigm remain to be investigated and confirmed in future studies.

\subsection{Relation to C-PAK (Correcting and completing variable-length Prefix-based Abbreviated Keys) paradigm} 
Our abbreviation-based writing scheme can be viewed as an implementation of the C-PAK (Correcting and completing variable-length Prefix-based Abbreviated Keys) text-input paradigm\cite{li2022c}. Compared to traditional forward predictions, C-PAK further exploits the redundancy and compressibility of text through the realization that the information that helps predicting a word is present not only in its preceding words (left context), but also in the words that follow it (right context), an insight shared with bidirectional neural LMs such as BERT\cite{devlin2018bert}. PhraseWriter, the first implementation of C-PAK, used count-based n-gram language models with narrow context windows. In a lab study, users of PhraseWriter showed lower average text-entry rate than a non-abbreviation-based Gboard baseline, which was attributable to the limited abbreviation-expansion accuracy and therefore frequent revisions through backspace and re-typing\cite{li2022c}.
To our knowledge,  this study demonstrated acceleration of text-entry in the C-PAK paradigm for the first time through the enhanced text entry rates, i.e., in the two eye-gaze users. The results in this study also push forward the pareto frontier of the KSR-accuracy tradeoff relation demonstrated by Adhikary et al.\cite{adhikary2021accelerating}, for both typing on mobile devices and with eye gaze. This acceleration was brought about by combining the greater context awareness of the LLM-based abbreviation expansion and a UI design that allows initial AE failures to be repaired without the costly use of delete or backspace keys. Featuring its affordances for spelling words and finding alternative words, the SpeakFaster UI allows users to expend minimal motor effort when the intended phrase couldn't be found with the initial abbreviation. It should be noted that the KSR achieved by the eye-gaze typists were in the range of 0.32 to 0.36. Despite significantly exceeding the KSR from their baseline typing paradigm, these KSRs still showed a significant gap from the offline simulation results (which exceeded 0.6). This was due to the unavoidable user errors when operating the SpeakFaster UI, such as mistakes and typos when entering the initials-only abbreviation and spelling out keywords, as well as mistakes when selecting phrase options from KeywordsAE and FillMask. These findings show that future work on improving the error tolerance of the LLMs is a fruitful future direction in further improving the speed of such abbreviation-based text entry UIs.

In addition to per-word abbreviations, previous studies have also explored expanding bags of keywords into full phrases using AI (e.g., KWickChat\cite{shen2022kwickchat}). Compared to KWickChat, which converts a bag of content words into full phrases, SpeakFaster opts to give users finer-grained control on exact sentence structure and wording. Two aspects of the SpeakFaster UI can be extended to bags of keywords. In particular, KWickChat can allow users to provide more keywords if the initial bag of keywords failed to yield the intended phrase; word replacement similar to FillMask also may also fit the KWickChat design. 

\subsection{Length of suggestions}
Can AE become a part of the conventional sequential typing, so that users can use AE when so desired and revert back to typing letter-by-letter otherwise? The answer to this question hinges on whether
sentence-level abbreviations consisting of word initials (e.g., "iipitb") can be reliably distinguished from non-abbreviated words. Another important factor that needs to be taken into consideration is the size of the screen. Small screens limit not just the number of options that can be surfaced, but also the length of the expanded text. This can introduce secondary effects wherein editing or correcting for errors in a smaller screen can be much more demanding and difficult. Thus sentence level expansion may be even less preferred.

\subsection{Tuning vs. prompting}
The current SpeakFaster system relies on two fine-tuned LLMs to reliably and accurately suggest text expansions. However, with remarkable improvements the field is seeing (e.g., GPT-4) it's conceivable that such expansions could simply be enabled by a single LLM and perhaps with no necessity for fine-tuning e.g., with prompt-tuning or even just good prompt engineering. This could make it much easier to develop SpeakFaster-like text-entry applications in the future.

\subsection{Latency and internet access}
Current LLMs are large in parameter count and the amount of computation required during inference, and thus can only be served on large dedicated servers. This necessitates the mobile and AAC devices to have access to the Internet. Further, with mobile text-entry applications, latency is an important consideration. As noted before, in our case latency to query the server will be a factor.  With recent advances in open source LLMs (e.g., LLAMA\cite{touvron2023llama}) and software for faster inference of LLM on local devices (e.g., GGML\cite{ggml2023}), in the near future, smaller more efficient models will be able to perform similarly.

\subsection{Cognitive overhead}
As seen in Fig.~\ref{fig:temporal_analyses} the current SpeakFaster UI leads to significant cognitive overhead for users, since they face burdens from multiple aspects of the task that do not exist in traditional sequential typing. This includes the mental conversion of the intended phrase into first-letter initialism, reviewing and selecting phrase and word options predicted by the LLMs, and deciding the next action (e.g., spelling vs. FillMask). While our abbreviation scheme is still simple, presence of contracted phrases (e.g., “isn't”, “we'll”) could take some time for users to adapt, and it would also be difficult to potentially correct abbreviations if the user didn't keep track of every word in a sentence. It remains to be seen though if things would change considerably with much longer term usage. However, the cognitive overhead appears to be within manageable bounds for our intended users, which is supported by the fact that all the users in our mobile user study and both eye-gaze users with ALS were able to learn the novel UI within the duration of the study sessions (see Supplementary Section~\ref{sec:supp_s5} for more details).

\subsection{Limitations and future directions}

Our KeywordAE LLM used in this study did not incorporate tolerance for typos in its input abbreviations1. Whether such tolerance benefits the usability and text-entry rate remains to be investigated in future studies. The SpeakFaster prototype reported in this study was created for English. However, the multilingual nature of today’s LLMs (e.g., LaMDA\cite{thoppilan2022lamda}, PaLM\cite{chowdhery2022palm}, and OPT\cite{zhang2022opt}) opens the door to extending the prototype to other languages. The extension to alphabet-based languages such as most European languages should be relatively straightforward and can be based on a similar UI design. Application on non-alphabet languages such as Chinese and languages that use mixed character sets such as Japanese are more complex and require additional considerations. However, it is worth noting that the “Jiǎnpīn” input paradigm, where the romanized form (Pinyin) of each Chinese character is abbreviated as one or more initial letters, is similar to the SpeakFaster UI design and formed an original inspiration for the more general C-PAK paradigm.

Several limitations of the current study remain to be improved upon by future studies. First, the SpeakFaster prototype we tested limited the phrases to ten words and mid-sentence punctuation or shorter. While this length limit is sufficient to capture a majority of sentences used in daily communication, the design must be extended to longer phrase length limits to support greater flexibility and expressiveness in the future. This extension requires improvements in both the underlying LLM and UI design, as multiple candidates of longer sentences cannot fit easily onto available screen area. Second, the acceleration achieved in this study was based on the assumption that the broader conversational context (words spoken by the eye-gaze user’s conversation partner) is available to the LLMs. The practicality of this context awareness for real-life AAC communication awaits further investigation integrating technical perspective\cite{adhikary2019investigating} and privacy considerations\cite{cai2023speakfaster}.

\section{Conclusion}
Typing letter-by-letter on an on-screen keyboard by eye-gaze tracking is a non-invasive means of communication used by people with severe motor impairments such as in ALS. The SpeakFaster project showed that the efficiency of such a practical paradigm of communication can be improved by an abbreviation-expansion (AE) system powered by two fine-tuned LLMs. In this system, the user first types the sentence initials which get expanded to five candidate sentences by the initials AE model. If the intended sentence is not found, the system enables two revision methods: incremental spelling of words (KeywordAE) and selecting from alternative words (FillMask). After confirming the usability of the system by employing able-bodied users on a mobile device, we obtained test results from eye-gaze typers with ALS showing that such users could indeed understand the UI design intentions, take advantage of the predictive power of the LLMs fine-tuned to the AE tasks, reduce the number of keystrokes entered by 14-46 absolute percentage points, and improve their typing speed by 29-60\%. Greater speed gain was observed in the eye-gaze users than the mobile users, which was attributable to the greater benefit of motor savings in the former group of users than the latter under a trade-off relation with increased  cognitive load of using the LLM-powered UI. The work shows that fine-tuned large language models integrated with UI design could bring real benefits to people with severe motor and speech impairments.

\section{Acknowledgements}
Mahima Pushkarna assisted with the design of the SpeakFaster UI.  Jon Campbell assisted with Windows programming and eye-tracker interfacing. William Ito and Anton Kast helped with UI prototyping and testing. Julie Cattiau, Pan-Pan Jiang, Rus Heywood, Richard Cave, James Stout, Jay Beavers, and John Costello helped through insightful discussions and other assistance. Shumin Zhai provided extensive feedback on the manuscript.

\bibliography{references}

\appendix
\newpage
\section{Supplementary Information}

\subsection{Fine-tuning of large language models}
\label{sec:supp_s1}
We used the 64-billion-parameter version of the LaMDA\cite{thoppilan2022lamda}, a pre-trained, decoder-only large language model (LLM) as the base model for fine-tuning on the KeywordAE and FillMask tasks in SpeakFaster. LaMDA was implemented in the Lingvo\cite{shen2019lingvo} framework which was in turn based on TensorFlow\cite{abadi2016tensorflow}. The base LaMDA consisted of 32 transformer1 layers (embedding and model dimensions: 8,192, feed-forward dimensions: 65,536, number of attention heads per layer: 128; activation function: ReLU) and was pre-trained for the next-token prediction objective on a large-scale text corpus containing 1.56 trillion tokens from a mixture of public web text and public dialogues\cite{thoppilan2022lamda}. LaMDA used a SentencePiece tokenizer\cite{kudo2018sentencepiece} with a vocabulary size of 32,000. To optimize serving efficiency (see Section~\ref{sec:supp_s3}), we fine-tuned two models for the KeywordAE and FillMask tasks separately.

The KeywordAE models were fine-tuned on data synthesized from four publicly-available dialogue datasets. These were the same dialogue datasets as used in Cai et al.\cite{cai2022context}. Each dialogue in the datasets consisted of a number of turns, yielding multiple examples for LLM fine-tuning and evaluation. Given a turn of a dialogue, we synthesized an example with three parts: Context, Shorthand, and the Full phrase. The context consisted of the previous turns of the dialogue, separated by the curly braces as delimiters. All turns of a dialogue except the first turn were associated with a non-empty context. The shorthand was an abbreviated form of the full text generated from SpeakFaster’s abbreviation rules. The three types of abbreviations that we included in the LLM fine-tuning are listed below.
\begin{enumerate}
    \item \textbf{Initials-only abbreviation:} Abbreviating each word as the initial letter in a case-insensitive fashion (e.g., “I saw him playing in the bedroom” $\xrightarrow{}$ “ishpitb”). Sentence-final punctuation was omitted, while sentence-middle punctuation (most commonly the comma) was preserved in the abbreviation (e.g., “ok, sounds good” $\xrightarrow{}$ “o,sd”). These aspects of the abbreviation scheme were identical to our previous report\cite{cai2022context}. However, to make the abbreviation rules easier to learn and memorize for users, the current study simplified the rule regarding contracted words, wherein a contracted word was abbreviated as only its initial (e.g., “you’re” $\xrightarrow{}$ “y” instead of “yr” as in the previous work\cite{cai2022context}). 
\item \textbf{Abbreviations with complete keywords:} These abbreviations contained one or more words completely spelled out, embedded in the initials-only abbreviation for the remaining words while respecting the original word order. For example, “I saw him playing in the bedroom” may be abbreviated as “ishpit bedroom”, where the last word “bedroom” is spelled out. The same phrase may also be abbreviated as “i saw hpit bedroom”, where both the 2nd and last words are spelled out. They enabled users to find phrases that couldn’t be expanded correctly based on the initials alone, which tend to be longer and less-predictable phrases.
\item \textbf{Abbreviations with incomplete keywords:} Similar to abbreviations with complete keywords, but allowing the keywords to be partially written. Compared to complete keywords, incomplete keywords afforded additional keystroke saving by further exploiting the redundancy and compressibility of written text. The fine-tuning of LaMDA incorporated two forms of incomplete keywords: prefix and consonant. In the prefix abbreviation scheme, a word can be abbreviated as a sequence of two or more characters that is a prefix of the word. For example, the word “bedroom” may be abbreviated as “be”, leading to the phrase abbreviation “ishpit be”. The consonant scheme omitted vowel letters (“a”, “e”, “i”, “o”, and “u”) if they were not at the beginning of a word and kept only the first few non-vowel letters, leading to a phrase abbreviation such as “ishpit bd” where “bd” is the shorthand for the word “bedroom”. The SpeakFaster UI tested in the user studies, however, only supported the prefix scheme for the sake of simplicity.
\end{enumerate}

To synthesize examples for abbreviations with complete keywords, we determined the number of keywords in a sentence by using a uniform distribution [1, 3], while ensuring that at least one word of the sentence is preserved as a non-keyword (the initial). To synthesize examples for abbreviations with incomplete keywords, we randomly selected the number of keywords from a sentence by using the [1, 5] uniform distribution, also ensuring that not all words of a sentence are selected as keywords. For a given word selected as the incomplete keywords, we limit the number of letters in its abbreviation to NL, with NL following a uniform distribution [2, 5]. An equal number of examples were synthesized for the prefix and consonant incomplete-keyword schemes. Table~\ref{tab:ft_data} summarizes the number of unique examples used in the fine-tuning of the KeywordAE LLMs. For KeywordAE v1, only the initials-only and complete-keyword examples were used; for KeywordAE v2, the dataset additionally included the incomplete-keyword examples, in order to support partially-spelled keywords. 

\begin{figure}
\centering
\includegraphics[width=\linewidth]{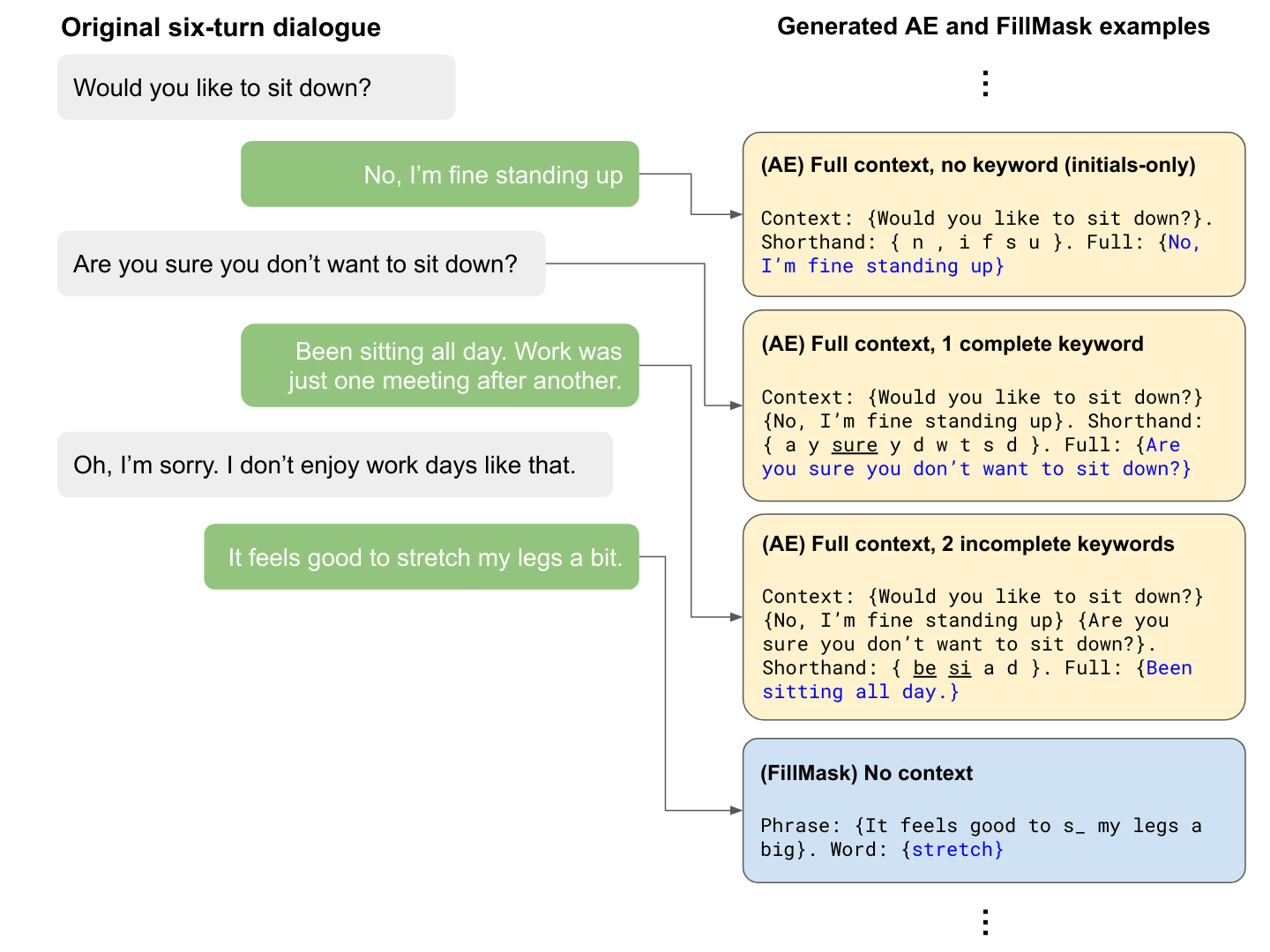}
\caption{Abbreviation schemes by examples. On the left we show a sample dialogue from the Turk Dialogues Corrected (TDC) dataset, which consists of six turns between two interlocutors. On the right we show instances of individual examples generated from the dialogue, used for training the KeywordAE and FillMask models. The three AE examples vary in their abbreviation scheme. The first one consists of an initials-only abbreviation, i.e., without any spelled keywords. The second one consists of a complete (fully-spelled) keyword. The third one contains two incomplete (partially-spelled) keywords. The “Shorthand” parts of the training examples contain the space character inserted between the letters and keywords that comprise the abbreviation, a trick employed to improve LaMDA fine-tuning accuracy by extracting more consistent tokenization from the underlying SentencePiece tokenizer\cite{kudo2018sentencepiece}. The underscores for the complete and incomplete keywords are added only in this figure for visual clarity. The FillMask example at the bottom shows an instance of no-context example. The actual dataset contains both no-context examples and examples with previous turns of the dialogues as context. In these examples shown, the text in black is the inputs to the LLMs, while the text in blue is the targets that the LLMs are trained to generate. The colors are added only for visual clarity of presentation in this figure.}
\label{fig:abbv_schemes}
\end{figure}

\begin{table}
    \centering
\resizebox{\columnwidth}{!}{
    \begin{tabular}{c|l|l|l|l}
\hline
\textbf{Dialogue}& \textbf{Abbreviation keyword} & \multicolumn{3}{c}{\textbf{\#(examples), tokens/example}} \\ \cline{3-5}
\textbf{Corpus} & \textbf{type or FillMask} & \textbf{Train} & \textbf{Dev} & \textbf{Test} \\ \hline
Turk Dialogues Corrected (TDC)\cite{vertanen2017towards} & (i) Initials-only & 8,590, 50.5$\pm$21.3 & 2,800, 50.6$\pm$21.4 & 2,800, 51.1$\pm$21.9 \\
 & (ii) Complete & 7,948, 51.5$\pm$21.4 & 2,569, 51.7$\pm$21.6 & 2,545, 52.1$\pm$21.9 \\
 & (iii) Incomplete & 15,158, 60.3$\pm$22.4 & 4,902, 60.3$\pm$22.9 & 4,898, 60.5$\pm$22.9 \\
 & (iv) FillMask & 8,584, 46.2$\pm$19.7 & 2,800, 46.5$\pm$20.0 & 2,798, 46.8$\pm$20.3 \\ \hline

Turk AAC\cite{vertanen2011imagination} & (i) Initials-only & 5,019, 21.5$\pm$5.3 & 559, 22.1$\pm$5.5 & 565, 21.0$\pm$5.0 \\
 & (ii) Complete & 4,352, 22.9$\pm$5.1 & 490, 23.3$\pm$5.3 & 488, 22.3$\pm$4.8 \\
 & (iii) Incomplete & 7,960, 29.9$\pm$8.1 & 904, 30.4$\pm$8.5 & 856, 29.0$\pm$7.9 \\
 & (iv) FillMask & 5,019, 19.4$\pm$2.8 & 559, 19.7$\pm$2.9 & 565, 19.2$\pm$2.6 \\ \hline

DailyDialog Corrected\cite{li2017dailydialog} & (i) Initials-only & 123,650, 86.1$\pm$70.4 & 9,233, 84.4$\pm$66.4 & 8,318, 82.8$\pm$64.0 \\
 & (ii) Complete & 99,202, 87.6$\pm$69.6 & 7,334, 85.9$\pm$64.9 & 6,612, 84.7$\pm$63.3 \\
 & (iii) Incomplete & 196,286, 95.6$\pm$701.3 & 14,588, 93.8$\pm$66.0 & 13,198, 92.4$\pm$63.7 \\
 & (iv) FillMask  & 123,426, 82.2$\pm$69.7 & 9,211, 80.4$\pm$65.7 & 8,306, 79.0$\pm$63.2 \\ \hline

Cornell Movie Dialogues\cite{danescu2011chameleons} & (i) Initials-only   & 327,569, 63.1$\pm$64.0 & 41,636, 60.7$\pm$60.3 & 36,772, 64.4$\pm$68.2 \\
& (ii) Complete    & 241,115, 67.4$\pm$65.1 & 30,268, 64.8$\pm$60.4 & 26,464, 68.8$\pm$68.0 \\
& (iii) Incomplete   & 472,746, 74.8$\pm$65.7 & 59,124, 72.3$\pm$61.2 & 52,380, 75.7$\pm$67.4 \\
& (iv) FillMask   & 323,834, 60.0$\pm$63.2 & 41,030, 57.8$\pm$59.5 & 36,057, 61.1$\pm$67.0 \\ \hline

\end{tabular}
}
\caption{Composition of the data for fine-tuning and evaluating LaMDA for the AE and FillMask and tasks. For each dialogue corpus, the statistics for four subsets are shown: (i) no keyword, i.e., initials-only abbreviations; (ii) Complete keywords: initials with complete keywords, with the number of keywords per sentence distributed uniformed from 1 through 3; (iii) Incomplete keywords: initials with incomplete keywords, of both the prefix and consonant schemes, with the number of incomplete keywords per sentence distributed uniformly from 1 through 3 and the length limit of incomplete keyword uniformly distributed from 2 through 5; (iv) FillMask examples. (i) - (ii) were used to fine-tune LaMDA for KeywordAE v1 (i) - (iii) were used in KeywordAE v2 fine-tuning; and (iv) was in FillMask fine-tuning. The tokens/example column shows the mean lengths of the examples and their standard deviations based on the SentencePiece tokenizer\cite{kudo2018sentencepiece} with 32,000 vocabulary items.}
    \label{tab:ft_data}
\end{table}

Fine-tuning of the AE model was performed on 32 Google Tensor Processing Unit (TPU) v3 chips connected in a 4x8 topology\cite{jouppi2017datacenter}. A maximum input length of 2,048 tokens was used. The training used a per-host batch size of 16 and a shared AdaFactor optimizer\cite{shazeer2018adafactor} under a fixed learning rate of $2\times10^{-5}$. The checkpoint for subsequent serving and evaluation was selected on the basis of minimal cross-entropy loss on the dev set, which occurred at fine-tuning step 12,600.

\begin{figure}
\centering
\includegraphics[width=\linewidth]{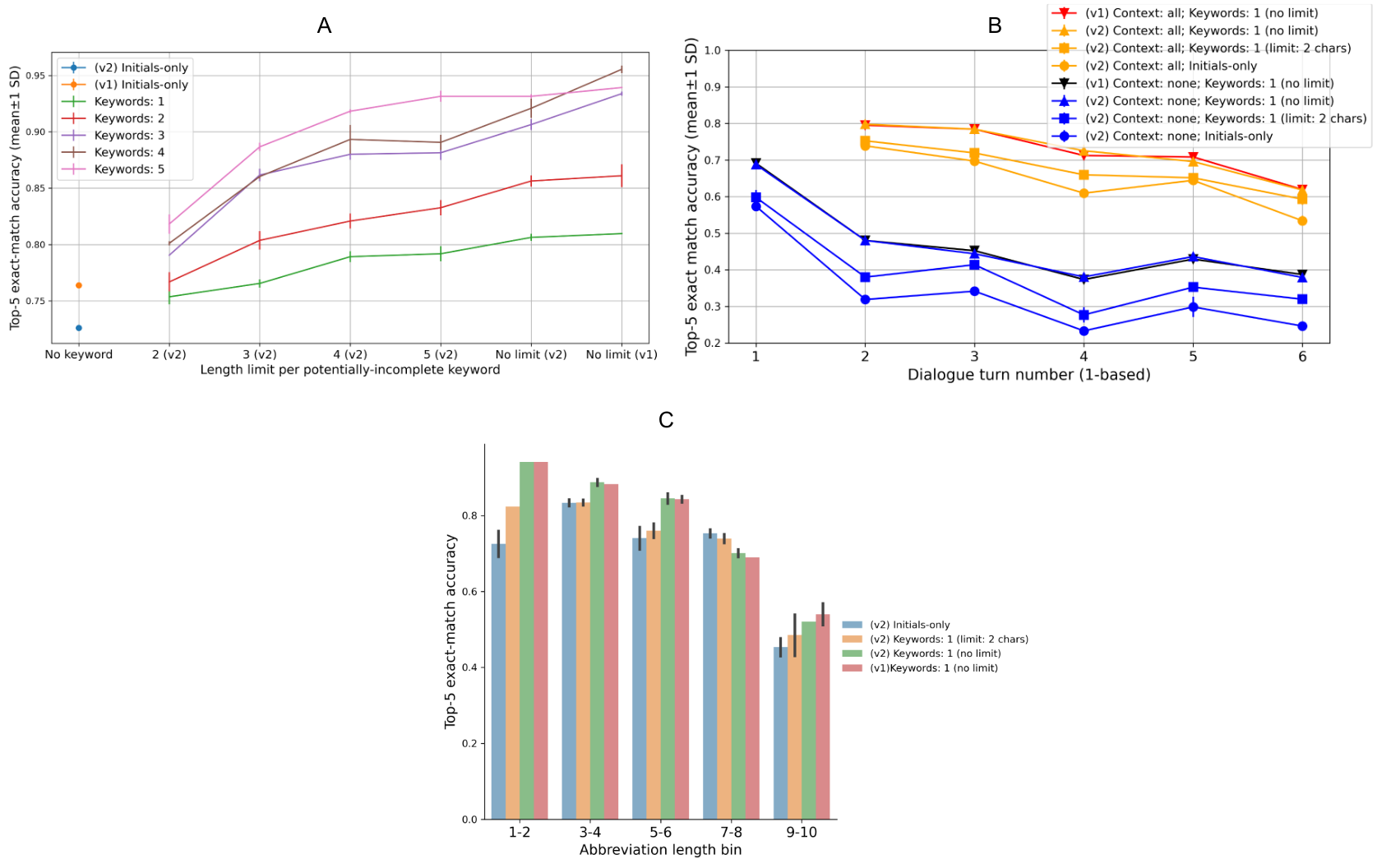}
\caption{Exact-match accuracy of the fine-tuned LaMDAs for abbreviation expansion, evaluated on the test split of the TDC corpus, under different levels of phrase and word-level abbreviation types and different amounts of dialogue context. The results are evaluated on the sentences from the test split of the Turk Dialogues Corrected (TDC) corpus for sentences of ten or fewer words and mid-sentence punctuation marks. The error bars who $\pm$1 SD around the mean for 3 repeated evaluation runs. The single data point on the left plots the exact-match AE accuracy with initials-only abbreviations (i.e., no keyword). \textbf{A:} The five curves show the AE accuracy when different numbers of keywords (from 1 to 5) are included in the abbreviation, with the keywords chosen at random. The x-axis is the limit on the length of each potentially-incomplete keyword, so that the rightmost two ticks (labeled “No limit (v2)” and “No limit (v1)”) correspond to using only complete keywords in the abbreviations. The complete-keyword accuracy of the dedicated v1 model slightly beat that of the v2 model. \textbf{B:} Representative relations between the accuracy of AE and dialogue turn number in a six-turn dialogue in the style of the TDC corpus, and how the relation is affected by the amount of dialogue context used when invoking the fine-tuned LaMDA for AE. For the orange curves, all previous dialogue turns were included as the context (e.g., when performing AE on dialogue turn \#4, the content of the first three turns were used as the context). In each set of curves, three curves are shown, for initials-only AE, incomplete keyword AE (limiting keyword length to two characters, and complete keyword AE). The sets of blue and orange curves show the v2 model’s accuracies without and with dialogue context, respectively. For comparison the accuracies of the v1 model for complete keywords are also shown (in black and red). \textbf{C:} AE accuracy as a function of phrase length as measured by the initials-only abbreviation length of a phrase (i.e., the number of words plus the number of mid-sentence punctuation). The first three different bar colors show the AE accuracies based on initials-only abbreviations, abbreviations with a single incomplete keyword limited to 2-character length, and a complete keyword, when used as input for the v2 model. The remaining bar color shows accuracies of the v1 model under complete keywords for comparison.}
\label{fig:accuracy_abbv_types}
\end{figure}

To quantify the accuracy of the fine-tuned LaMDA on the AE task with the three types of abbreviation mentioned above, we evaluated the exact-match accuracy among top-5 predictions. For each AE input, 128 outputs were sampled from the temperature-based sampling if the total number of initials and keywords in the abbreviation was $\leq$5, else it was 256. The exact match was determined after text normalization including whitespace standardization, folding to lowercase, and omitting sentence-final punctuation. The sampling was based on a temperature of 1.0, which was determined on the Dev split to be the optimal sampling temperature value for AE. Panel A of Fig.~\ref{fig:accuracy_abbv_types} shows the evaluation results on the 2nd turn of the dialogues by using the first turn as the context. The general trends of relation between AE accuracy with abbreviation type and length is similar for other turns of the dialogues and when no dialogue context was used. On average, with initials-only abbreviation, 72.6\% and 76.3\% of the sentences could be expanded into the exact-matching phrase (the single data point labeled blue in Fig.~\ref{fig:accuracy_abbv_types}, Panel A) based on the v1 and v2 models, respectively. As expected and shown by the curves in Fig.~\ref{fig:accuracy_abbv_types}A, incorporating keywords (both complete or incomplete) in the abbreviation progressively increased the percentage of successful expansion. The amount of improvement shows a monotonic relation with both the number of keywords and the number of characters used to represent each keyword. The v1 model showed slightly better accuracy over the v2 model for the initials-only and complete-keyword cases, which may be due to the v1 model’s specialization on the more limited input format that doesn't include partially-spelled keywords.

To quantitatively illustrate the benefits of dialogue context, a similar evaluation was performed on all six turns of the TDC dialogues (test split), with and without dialogue context. The accuracies are plotted as a function of dialogue turn number in Panel B of Fig.~\ref{fig:accuracy_abbv_types}. Comparing the sets of blue and orange curves, we see a clear and substantial boost to the v2 KeywordAE model’s accuracy due to the dialogue context, which reflects the reduction in the space of possible phrases (given abbreviation) resulting from contextualizing on the previous turns of a dialogue. The panel also shows that this boost of dialogue context is seen for all three styles of abbreviations: initials-only, incomplete keywords, and complete keywords. The latter two show additional accuracy enhancement on top of the contextual boost, reflecting the expected trend that the fine-tuned LaMDA model predicts phrases more accurately when more characters become available, regardless of whether these additional characters form complete words. The effect of context and dialogue turns on the v1 KeywordAE model (the black and red curves in Fig.~\ref{fig:accuracy_abbv_types}B), which doesn't not support incomplete keywords, closely resembles those on the v2 model. 

Panel C of Fig.~\ref{fig:accuracy_abbv_types} shows the relation between AE accuracy and length of the phrase. We see a general decreasing trend for the accuracy of AE with increasing phrase length for initials-only AE, as well as incomplete and complete keywords. We can also see a strong trend in which augmenting the abbreviations with more characters, in the form of incomplete or complete keywords, boosts accuracy for all phrase lengths. The relations between AE accuracy and phrase length are similar between the v1 and v2 KeywordAE models.

These results form the basis for designing the SpeakFaster UI (Fig.~\ref{fig:pathways}) that allows users to amend abbreviations that fail AE by using more and more characters. The fact that even with five fully-spelled keywords, only 93.1\% of the phrases could be expanded correctly is one of the motivations for allowing the user to use the additional remedying approach of FillMask and the ever-existing option of spelling out the words in full in the SpeakFaster UI.

To fine-tune LaMDA on the FillMask task, we use the same four public dialogue datasets as above. However, instead of including abbreviated forms of the phrases, we synthesized examples of a phrase with a single word masked, followed by the ground truth word. Therefore, instead of triplets of context-shorthand-full in the AE fine-tuning, the FillMask fine-tuning was based on context-phrase-word triplets such as (see also bottom right of Fig.~\ref{fig:abbv_schemes}):

\textbf{Context}: \{Been sitting all day. Work was just one meeting after another.\} \textbf{Phrase}: \{Oh, I'm s\_.\} \ \textbf{Word}: \{sorry\}

In the example above, “sorry” is the masked word. The “Phrase” part contains the masked word with its initial letter embedded in the context words in the correct order. These examples for FillMask avoided masking words that start with a number or a punctuation mark. Based on evaluation on the Dev split, we determined the optimal sampling temperature of the FillMask to be 2.0, which was higher than the optimal sampling temperature for KeywordAE (1.0).

\begin{figure}[t]
\centering
\includegraphics[width=0.55\linewidth]{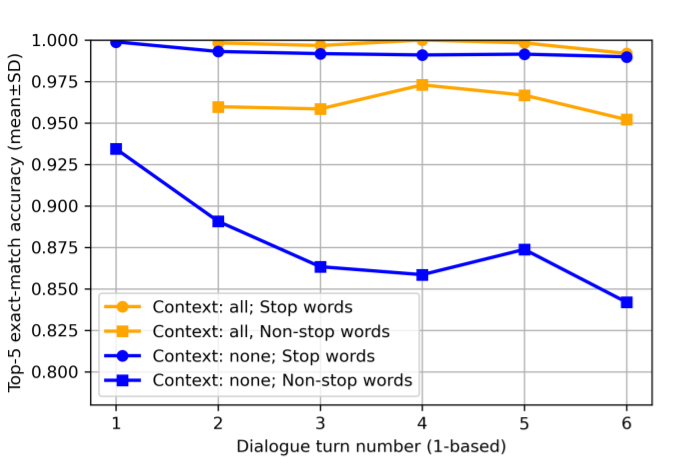}
\caption{Exact-match accuracy of the fine-tuned LaMDA for FillMask task, evaluated on the test split of the TDC corpus, under different amounts of dialogue context for predicting masked stop words and non-stop words. The results are evaluated on the same sentences used for the evaluation of the fine-tuned LaMDA for abbreviation expansion. The results are plotted similarly to Figure~\ref{fig:accuracy_abbv_types} Panel B. In each set of blue or orange curves, a curve is shown for the accuracy of predicting masked stop words and non-stop words, respectively.}
\label{fig:fillmask_eval}
\end{figure}

Figure~\ref{fig:fillmask_eval} shows the results of evaluating the accuracy of the fine-tuned FillMask LaMDA model on sentences with randomly selected and masked words from the test split of the TDC dataset. Stop words (i.e., high-frequency words such as “the” and “for”) have significantly higher prediction accuracy than non-stop words. As in the AE evaluation results (cf. Fig.~\ref{fig:accuracy_abbv_types}, Panels A - B), incorporating conversational context increased the FillMask accuracy for both the stop and non-stop word categories. However, the benefit of context was much more pronounced for non-stop words (7 - 11 percentage points) than for stop words ($\leq$1 percentage points).

\subsection{Detailed offline simulation results}
\label{sec:supp_s2}

\begin{figure}[t]
\centering
\includegraphics[width=0.55\linewidth]{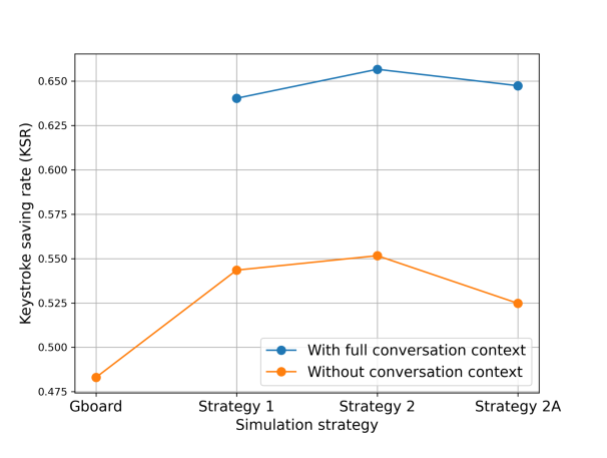}
\caption{Comparing the keystroke saving rate (KSR) from simulations ran on three different use strategies of SpeakFaster as well as the Gboard baseline. The results shown in this figure are based on presenting top-5 options from AE, FillMask, and Gboard word suggestions. The blue curve shows the KSR values obtained from full utilization of conversation context by the AE and FillMask models of SpeakFaster, i.e., all the previous turns of a dialogue. It does not contain data for Gboard because Gboard’s n-gram model does not utilize the previous dialogue turns. The orange curve shows the results obtained from not utilizing the conversation context.}
\label{fig:ksr_vs_simulation}
\end{figure}

The simulation results shown in the main article were based on utilizing all available conversational context. Specifically, when performing AE or FillMask on the n-th turn of a dialogue in the TDC corpus (test split) that we use for simulation, the 1st through (n-1)-th turns of the dialogues are fed to the fine-tuned LLMs to provide them with the maximal context for performing their respective inference tasks. However, in real-life situations, due to the constraints related to privacy, system capabilities, and the modality of text and speech communication, the contextual information may not always be available or accurate. The orange curve in Figure~\ref{fig:ksr_vs_simulation} shows the KSR when no conversational context is used for inference with the fine-tuned LLMs, which show substantial decreases (10 percentage points or more) compared to the corresponding KSRs obtained under full context. This pattern is seen for all three simulation strategies (1, 2, and 2A), which were based on different degrees of utilization of the keyword AE and FillMask features. This observation corroborates the aforementioned evaluation results on the AE and FillMask models in emphasizing the benefit of wider context in LLM-based communication UIs. However, note that SpeakFaster provided higher KSRs than Gboard’s forward word prediction even when no wider conversational context was available.

\subsection{Serving of large language models}
\label{sec:supp_s3}

During the simulations and user studies, for both KeywordAE and FillMask, the fine-tuned LLMs were served with 16 Google TPU v3 chips[5] in a 4x4 configuration. To increase inference speed, we performed post-training quantization to the bfloat16 dtype15 on both the KeywordAE and FillMask LaMDA models. The served KeywordAE and FillMask models were configured to a maximum input length of 128 SentencePiece tokens4, which was sufficient to accommodate up to five contextual dialogue turns, each consisting of ten or fewer words for all the test dialogues used in the lab study with LP1. SentencePiece tokens are variable length tokens learned from the training corpus. Short and frequent words such as “the” and “time” are single tokens, while longer and less frequent tokens such as “understanding” are split as two or more tokens. The AE model was served with a maximum decoder step of 20 tokens, which is sufficient to capture the vast majority of phrases with ten or fewer words and mid-sentence punctuation. To serve the FillMask model, however, we exploited the fact that the output of the model is a single word and hence could use a smaller value for maximum decoder steps (6) to achieve a shorter serving latency.

\subsection{User study design and details}
\label{sec:supp_s4}

The mobile and eye-gaze user studies were based on the same user-interface code base, available in open source at \url{https://github.com/TeamGleason/SpeakFaster/tree/main/webui}. 

The mobile user study consisted of 19 participants. Ten participants were recruited from employees of Google working at its Cambridge, MA office; the remaining nine participants were recruited from the Boston area. They were all adults with associate-level education or higher and were speakers of American English.  During a user-study session, the user was comfortably seated, facing a Samsung Galaxy Tab S6 10.4-inch tablet placed in a stable level position on a tabletop. The device remained in landscape orientation. It presented the SpeakFaster UI running as an app at the top half of the screen and Gboard at the bottom half. Both SpeakFaster and Gboard were displayed along the full width of the tablet screen.

\begin{figure}[t]
\centering
\includegraphics[width=0.65\linewidth,height=10cm]{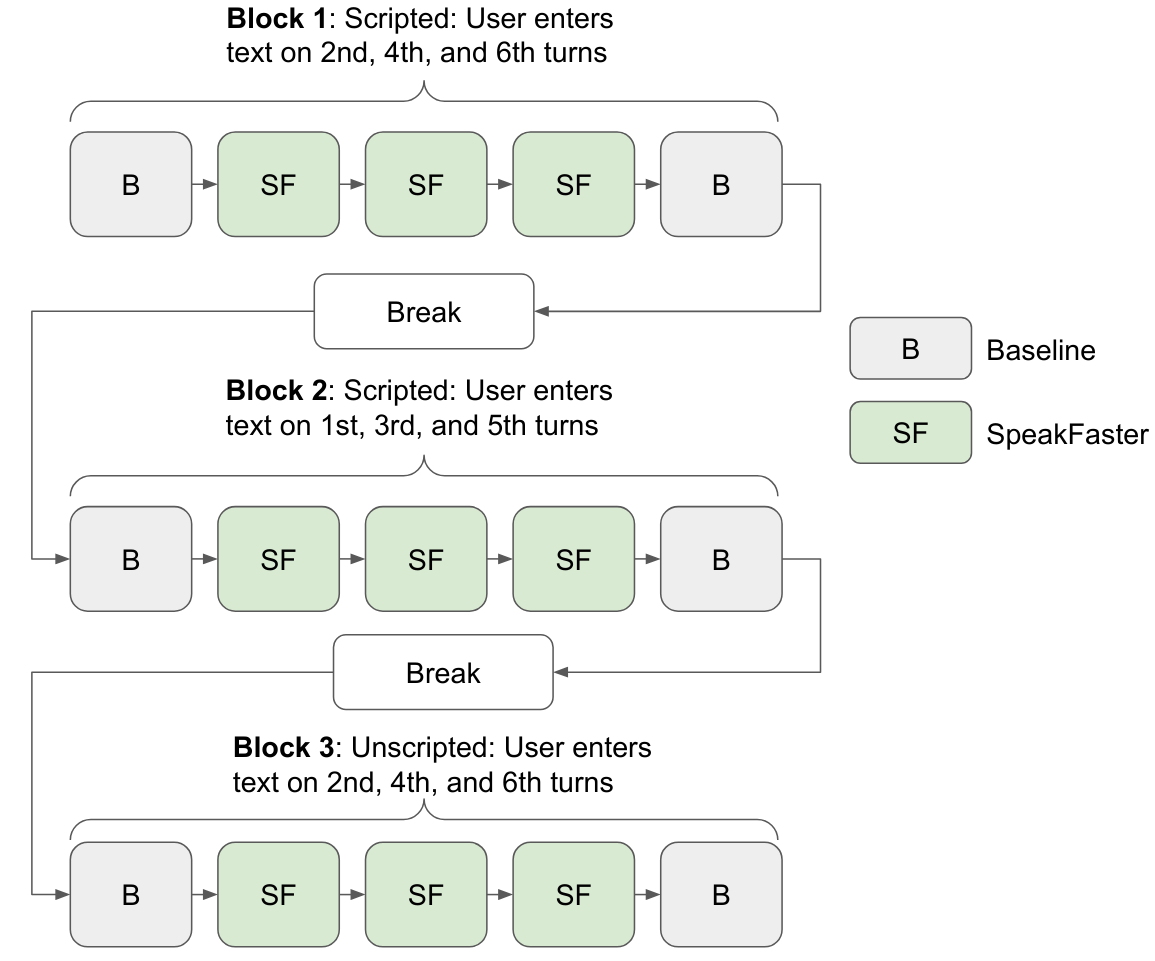}
\caption{Design of the lab study with participant LP1. Each box shows a six-turn dialogue used for testing. The gray and green boxes shows trials of the baseline condition and the SpeakFaster condition, respectively.}
\label{fig:lab_study_design}
\end{figure}

In the lab studies (including the mobile user study sessions and the session with the eye-gaze lab study participant), the ten scripted dialogues were arranged in two blocks. During the first block of five dialogues, the user played the role of the second interlocutor and entered the text for the 2nd, 4th, and 6th turns. In the second block of five dialogues, the user played the role of the first interlocutor and entered the text for the 1st, 3rd, and 5th turns. The ordering of the ten dialogs used in the first two blocks were randomized on a user-by-user basis. Following the scripted phase was an unscripted phase to measure the performance in a spontaneous text entry task in which the user typed their intended phrases instead of prescribed ones. 

During the lab study, the experimenter operated a companion web application on a laptop to initiate test dialogues. A scripted dialogues proceed without experimenter intervention once started. An unscripted dialogue required the experimenter to supply responses in the 3rd and 5th turns. The experimenter used the same companion application to send those dialogue turns.

These ten dialogues are selected from the test split of the Turk Dialogues dataset11, hence were not seen by the LLMs during their fine-tuning for KeywordAE or FillMask. We selected the dialogues based on the criteria of: 1) consisting of exactly one sentence per dialogue turn, 2) each sentence contains ten or fewer words, and 3) contain no potentially offensive or inappropriate content. The IDs of the ten dialogues selected dialogues were: chain2\_633\_298\_601\_4\_948\_1016, 
chain2\_431\_1021\_754\_865\_533\_290, 
chain3\_242\_108\_214\_13\_28\_50, 
chain2\_450\_34\_1143\_868\_23\_220,\\
chain2\_2121\_755\_43\_246\_381\_709,  
chain2\_752\_401\_1278\_363\_1136\_1091,  
chain2\_1578\_1057\_1229\_143\_692\_855, \\ 
chain2\_58\_150\_868\_489\_383\_264,
chain2\_58\_150\_868\_489\_383\_264, and 
chain3\_112\_258\_148\_288\_146\_242, which can be cross reference with the raw dataset at \url{https://aactext.org/turk/turk/turk-dialogues.txt}. In the dialogue with ID chain3\_242\_108\_214\_13\_28\_50, we fixed a grammatical error in the 5th turn, ``I’ve tried too'' $\xrightarrow{}$ ``I’ve tried to''. Eight additional dialogues that satisfy these criteria were selected for practice purposes. Their IDs were 
chain2\_230\_210\_1152\_921\_190\_391,  
chain2\_1854\_397\_1176\_1052\_141\_408,  
chain3\_161\_342\_288\_171\_313\_222, \\
chain2\_888\_773\_852\_406\_924\_605, 
chain3\_45\_147\_301\_192\_331\_261, 
chain2\_1481\_1455\_10\_53\_610\_18, \\ 
chain2\_1052\_582\_996\_1165\_267\_929,  
chain2\_839\_1478\_209\_43\_1020\_495.

The unscripted dialogues were prompted with the following starter questions:
\begin{itemize}
 \item "How do you like the weather today?"
 \item "What did you do yesterday?"
 \item "What pets have you had?"
 \item "What kind of books do you read?"
 \item "What kind of music do you listen to?"
 \item "What sports do you like to watch?"
 \item "Where would you like to go for vacation?"
 \item "Do you like the city you live in?"
\end{itemize}

The lab study participant was instructed to respond to these questions with dialogue turns that each consisted of a single sentence with ten or fewer words. Each unscripted dialogues lasted six turns (i.e., the same as the scripted ones), in which the participant was the interlocutor in the 2nd, 4th, and 6th turns, while the experimenter performed the 3rd and 5th turns impromptu.

Figure~\ref{fig:lab_study_design} shows the protocol of the lab study with participant LP1. It was arranged in three blocks, with interspersed breaks. Each block contained five six-turn dialogues, the first and last of which were the baseline condition of using the Tobii eye-gaze keyboard, while the three dialogues in the middle were based on the SpeakFaster UI. The first and second blocks used scripted dialogues from the test split of the TDC corpus. They differed in whether the participant played the role of the first or second interlocutor. The third block was based on unscripted dialogues in which the participant responded in the 2nd turn to the starter questions listed above, in addition to following up with the experiment in the 4th and 6th turns.

For lab study with participant LP1, the user’s word prediction history of the Tobii Windows Control virtual keyboard was cleared immediately before the start of the experiment, in order to prevent personal usage history of the soft keyboard from interfering with the testing results.

\begin{figure}[t]
\centering
\includegraphics[width=\linewidth]{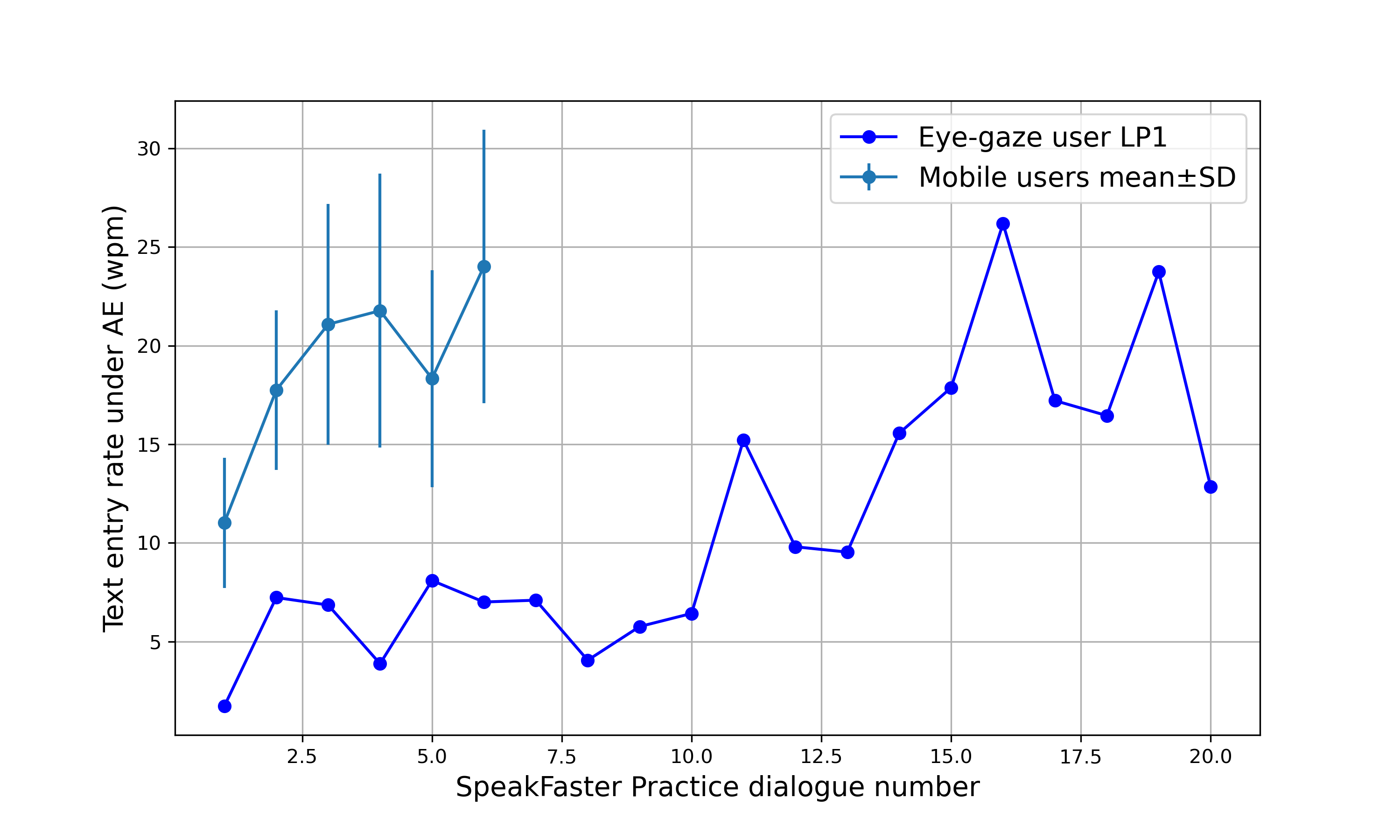}
\caption{ Learning curve of eye-gaze user LP1 on the SpeakFaster text-entry paradigm over 20 practice dialogues, in comparison with the average learning curve of the 19 users in the mobile study. In LP1, the practice was conducted over two separate days before the lab study. }
\label{fig:learning_curve}
\end{figure}

Figure~\ref{fig:learning_curve} shows how the lab study user LP1 gradually improved in the efficiency of text entry in the SpeakFaster UI through initial practice prior to the actual lab study over two days, in comparison with the same learning curve from averaging the 19 mobile lab-study participants. It can be seen that despite showing a slower learning rate, approximately ten practice dialogues were sufficient to train LP1 beyond the initial slow rate (< 5 WPM) and reach a level > 10 WPM. This shows that for eye-gaze communication users with cognitive abilities within normal limits at least, SpeakFaster has a manageable learning curve.

\subsection{Accuracy and speed trade-off}
\label{sec:supp_s5}

Adopting SpeakFaster's abbreviation expansion resulted in significant speed ups, but in some cases it reduced accuracy. For the lab study user LP1, a third (six) of the 18 scripted dialogue turns performed with the SpeakFaster UI contained errors, i.e., mismatches between the entered sentence and the sentence prompt. This led to an average word error rate (WER) of 15.4\% across LP1’s SpeakFaster turns, which was higher than the word error rate observed on the same participant under the 12 dialogue turns of baseline (non-SpeakFaster) condition: 0\%. However, as determined by the speech-language pathologists on our research team, only two of the errors were categorized “major”, i.e., affects the meaning of comprehensibility of the text. The remaining four errors were categorized as "minor" errors that did not affect the meaning of the sentence or its comprehension in a significant way, but may have resulted in incorrect grammar or idiom. This indicates that a majority of the word errors seen in the SpeakFaster-entered phrases did not hinder the communication in terms of intentions and meaning, although it may still impact the perception of the user’s communication or cognitive abilities in undesirable ways. Future studies can explore automatic correction\cite{bryant2023grammatical} of such errors by leveraging LLMs and context awareness.

\subsection{Temporal, Cost, and Error Analyses of Text Entry}
\label{sec:supp_s6}

When analyzing the results from offline simulations and event logs from user studies, we used the definition of keystroke saving rate (KSR):

\begin{equation}
    KSR = 1 - \frac{N_a}{N_c}
\end{equation} 

where $N_c$ is the character length of the entered phrase and $N_a$ is the number of total keystrokes and UI actions that the user performs to enter the phrase. The keystrokes include the ones used to enter the abbreviation, the ones used to specify keywords (complete or incomplete), and the ones used to type the words that cannot be found with FillMask. The UI interactions include the clicks used to enter the Spelling or FillMask mode, the clicks used to specify which word to spell, the clicks that specify which word to get replacements for through FillMask, in addition to the clicks used to select matching results from KeywordAE and FillMask. The actions that select the final phrase for text-to-speech output (e.g., gaze-clicking the speaker buttons in Fig.~\ref{fig:pathways}) are not included in the KSR calculation.


\end{document}